\documentclass[structabstract]{aa}  

\usepackage{graphicx}
\usepackage{txfonts}
\usepackage{natbib}
\usepackage{longtable}
\begin{document}
\title{A line confusion limited millimeter survey of \object{Orion
 KL}\\ II. Silicon-bearing species\thanks{Appendix A (density diagnostic), Appendix B (online Figures), and Appendix C (online Tables)
     are only available in
     electronic form via http://www.edpscience.org}\fnmsep\thanks{This work was based on observations carried out with the IRAM 30-meter telescope. 
IRAM is supported by INSU/CNRS (France), MPG (Germany) and IGN (Spain)}}

\author{Bel\'en Tercero\inst{1}
 \and Lucie Vincent\inst{1,2} 
  \and Jos\'e Cernicharo\inst{1} 
  \and Serena Viti\inst{3}
  \and N\'uria Marcelino\inst{1,4}
  }
   \institute{Centro de Astrobiolog\'ia (CSIC-INTA). Departamento de
              Astrof\'isica Molecular. Ctra. de Aljalvir Km 4, 28850
              Torrej\'on de Ardoz, Madrid, Spain.
    \and LERMA and UMR 8112 of CNRS, Observatoire de Paris-Meudon, 92195 Meudon Cedex, France.
    \and Department of Physics and Astronomy, University College London, Gower Street, WC1E 6BT, 
London, UK.
    \and National Radio Astronomy Observatory, 520 Edgemont Road, Charlottesville, VA 22903, USA.
    \\
              \email{terceromb@cab.inta-csic.es; lucie@damir.iem.csic.es; 
jcernicharo@cab.inta-csic.es; sv@star.ucl.ac.uk; nmarceli@nrao.edu}
}

   \date{Received September 29, 2010; accepted December 3, 2010}

  \abstract
{We present a study of the Silicon-bearing species detected
in a line confusion limited survey towards Orion KL performed 
with the IRAM 30-m telescope.}
{The analysis of the line survey is organized by families of molecules. Our aim is to 
derive physical and chemical conditions for each family taking into account all observed
lines from all isotopologues of each species.
Due to the large number of transitions in different 
vibrationally excited states covered by our data,
which range from 80 to 280 GHz, we can provide reliable 
source average column densities (and therefore, isotopologue
abundances and vibrational temperatures) for the detected molecules.
In addition, we provide a wide study of the physical properties
of the source based on the different spectral components found
in the emission lines.}
{We have modeled the lines of the detected molecules
using a radiative transfer code, which permit us to choose 
between Large Velocity Gradient (LVG) and Local Thermodinamic
Equilibrium (LTE) approximations depending on the physical conditions
of the gas. We have used appropriate 
collisional rates for the LVG calculations. In order to qualitatively 
investigate the origin of the SiS and SiO emissions 
in Orion KL we ran a grid of chemical models.
}
{For the $\textit{v}$=1 state of SiO we have detected the $J$=2-1
line and, for the first time in this source, emission in the 
$J$=4-3 transition, both
of them showing strong masering effect.
For SiO $v$=0, we have detected $^{28}$SiO, $^{29}$SiO, and $^{30}$SiO;
in addition, we have mapped the $J$ = 5-4 SiO line.
For SiS, we have detected the main species, $^{29}$SiS,
and SiS $\textit{v}$=1. 
Unlikely other species detected in Orion KL (\object{IRc2}), the emission peak of SiS 
appears at a velocity of $\simeq$15.5 km s$^{-1}$; a study of the 5-4
SiO line around IRc2
shows this feature as an extended component that
probably arises from the interaction of the outflow with the ambient cloud.
We derive a SiO/SiS column density ratio of $\simeq$13 
in the plateau component, four times lower than the
cosmic O/S ratio $\simeq$ 48. 
In addition, we provide upper limits to the column density
of several non-detected Silicon-bearing species. 
The results of our chemical models show that while it is possible
to reproduce SiO in the gas phase (as well as
on the grains), SiS is a product of surface reactions,
most likely involving direct reactions of Sulphur with Silicon.
}
{}


\titlerunning{Survey towards Orion KL: Silicon bearing species}  
\authorrunning{B. Tercero et al.}

\maketitle
%

\section{Introduction}
\label{sect_int}
The Orion BN/KL \citep{Bec67, Kle67}
nebula is one of the most studied star formation 
regions in the Milky Way.
At a distance of 414 pc \citep{Men07} the nebula is embedded in 
a giant molecular cloud
harboring practically all phases of the interstellar medium, 
from hot and diluted
plasma, to PDRs, protostellar cores, molecular outflows, SiO 
and H$_2$O masering regions,
high density cores, intermediate and high mass star 
formation, protoplanetary disks, and proplydes 
(see, e.g., \citealp{Gen80, Gen89, Wri95, Cer90, Cer94, Pla09}). 

Together with Sgr B2, Orion
BN/KL nebula exhibits a rich spectrum 
(see, e.g., Tercero et al., 2010, hereafter Paper I, and references
therein) produced by complex organic molecules which
are formed through reactions on the
grain surfaces during the collapse phase followed by evaporation 
when radiation from a newly
formed star becomes available. Due to the high 
temperature of the gas, molecular lines are particularly 
strong in Orion, allowing several line surveys of this source over the last 20 years.
Recently, we have performed a line survey towards Orion IRc2 source 
between 80 and 280 GHz 
(Paper I), not limited by sensitivity but only by line
confusion. The data
provide a significant number of transitions for all molecules 
detected so far towards this source. 
Although physical structure of the Orion is rather complex, 
the large number of transitions
observed for each species allows to model the different cloud 
components and to derive 
reliable physical parameters. In addition, the line survey 
provides a deep insight on the 
chemistry of the Orion KL region and allows to refine our knowledge of its 
chemical structure by searching for new molecular species
and new isotopologues and vibrationally excited 
states of molecules already known to exist in this source. 
In a first paper we have presented the line survey and 
analyzed the CS-bearing species 
deriving the abundance of CS, OCS, CCS, CCCS, H$_2$CS, and HCS$^+$
(Paper I). 
In this paper we 
will analyze the Silicon-bearing species SiO and SiS. 
SiO has been observed with single
dishes and interferometers (see, e.g., \citealp{Pla09} and references
therein), 
while only a few observations
are available for SiS
\citep{Dic81, Ziu88, Ziu91}.

SiO lines in Orion show a complex pattern of thermal and maser
emission. The masers seem to arise
from a small region around the radio continuum \object{source $I$} 
\citep{Chu87}, a young star 
with a very high luminosity without infrared
counterpart, $\simeq$10$^5$ L$_\odot$,  
\citep{Gez98, Gre04}. This source is also driving the low
velocity outflow observed in SiO
\citep{Beu05, Pla09} and in other molecular species (Paper I).
Recent studies have discussed the driving source of the high 
velocity outflow: whereas \citet{Beu08} claimed that \object{SMA1} (a 
sub-millimeter source not detected at IR and centimeter wavelengths,
predicted by \citealt{deV02} and detected by \citealt{Beu04})
is the host of the high velocity outflow (based on combined observations
of $J$=2-1 C$^{18}$O from the SMA and the 
IRAM 30m telescope), \citet{Pla09}
defended that this outflow is a continuation of the low velocity outflow
(based on CARMA observations of SiO {\it{v}}=0 $J$=2-1).

The SiO $v$=1 maser emission was modeled in
early interferometric observations as arising from a rotating 
and expanding disk \citep{Pla90}. However,
maser emission was also found in the $v$=0 $J$=2-1 line by \citet{Wri95} 
adding more complexity to the modeling
of the structure of the emitting source.
Recently, \citet{God09} have found maser emission
in the $v$=0 $J$=1-0 transition of $^{29}$SiO and $^{30}$SiO associated
with source $I$. In addition, observations 
of \citet{Pla09} clearly show that the
emission of this line arise essentially from an outflow driven by
source $I$. Although our single dish data
cannot provide a view of the spatial structure of the 
thermal and maser emission around that source, the
observed maser and thermal lines can provide useful 
constraint on the physical conditions of the gas.

In addition to the study of 
SiO and SiS we derive upper 
limits to the abundance of SiC, SiC$_2$,
c-SiC$_3$, SiC$_4$, SiN, SiCN, SiNC, ob-SiC$_3$, l-SiC$_3$, 
Si$_3$, SiCCO, SiCCS, SiH$_2$, H$_2$CSi, and the different isomers
of Si$_2$H$_2$. The observations are described in Sect. \ref{sect_obs}. 
The results for SiO and SiS are analyzed
in Sect. \ref{sect_res}. Section \ref{sect_phy} is devoted to 
the modeling of the observed lines. All
these results are discussed in Sect. \ref{sect_dis} in terms of comparisons 
with chemical models predictions
for Silicon-bearing species. The effect of the new collisional rates for 
SiO and SiS is analyzed in the Appendix. 

\section{Observations and Data Analysis}
\label{sect_obs}

\begin{table}
\begin{center}
\caption{$\eta_{MB}$ and HPBW along the covered frequency range\label{tab_eta_hpbw}}
\begin{tabular}{lll}
\hline 
\hline
 Frequency (GHz) & $\eta_{MB}$ & HPBW ('')\\
\hline
86 & 0.82 & 29.0\\
110 & 0.79 & 22.0\\
145 & 0.74 & 17.0\\
170 & 0.70 & 14.5\\
210 & 0.62 & 12.0\\
235 & 0.57 & 10.5\\
260 & 0.52 & 9.5\\
279 & 0.48 & 9.0\\
\hline
\end{tabular}
\end{center}
\tablefoot{
$\eta_{MB}$ and HPBW along the covered frequency range.
}
\end{table}

The observations were carried out using the IRAM 30m radiotelescope
during 2004 September (3 mm and 1.3 mm windows), 2005 March 
(full 2 mm window), 
2005 April (completion of 3 mm and 1.3 mm windows), 2007 January (maps and  
different positions) and March 2008 (2-D line survey using a multibeam
receiver). Four SIS receivers
operating at 3, 2, and 1.3 mm were used 
simultaneously with image sideband rejections within 20-27 dB 
(3 mm receivers), 
12-16 dB (2 mm receivers) and 13 dB (1.3 mm receivers). 
System temperatures were 
100-350 K for the 3 mm receivers, 200-500 K for the 2 mm 
receivers and 200-800 K for the 1.3 mm receivers, depending on the particular 
frequency, weather conditions, and source elevation. For the
spectra between 172-178 GHz, the system temperature was significantly
higher, 1000-4000 K, due to proximity of the atmospheric water line at 
183.3 GHz. 
The intensity scale was calibrated using two absorbers at
different temperatures and using the Atmospheric Transmission Model 
(ATM, \citealp{Cer85}; \citealp{Par01a}). 
The half power beam width (HPBW) and the main beam efficiency ($\eta_{MB}$)
along the covered frequency range are given in Table \ref{tab_eta_hpbw}.

Pointing and focus were regularly
checked on the nearby quasars 0420-014 and 0528+134. The observations
were made in the balanced wobbler-switching mode, with a wobbling frequency 
of 0.5 Hz and a beam throw in azimuth of $\pm$240''.
No contamination from the off position affected our observations 
except for a marginal one at the lowest elevations ($\sim$ 25$^{\circ}$) for 
molecules having emission along the extended ridge.

The backends used were two
filter banks with 512$\times$1 MHz channels and a 
correlator providing two 512 MHz bandwidths and
1.25 MHz resolution. We pointed
towards the (survey) position $\alpha$ = 5$^h$ 35$^m$
14.5$^s$, $\delta$ = -5$^{\circ}$ 22' 30.0'' (J2000.0) corresponding to
IRc2. 

The spectra shown in the figures are in units of main beam
temperature, $T_{MB}$. In spite of the
good image band rejection of the receivers, each setting was repeated 
at a slightly shifted frequency (10-20 MHz) in order
to identify and remove all features arising from the image side band. 
In the data reduction we have removed most of them above 0.05 K (see
Paper I for further explanation of this procedure).
\subsection{2D survey observations}
The HERA multibeam receiver was used to carry out 
a systematic line survey between 216--250 GHz over a region of 144x144'' 
(2x2' approximately) with a 4'' spacing.
System temperatures were around 400--500 K.
The sensitivity of the resulting
maps was comparable to those of the pointed line survey with the single
pixel SIS receivers.
The 2D line survey observations were performed in {\it On-The-Fly} mapping mode, with position 
switching using a emission-free reference position at an offset (-600'',0) from IRc2.
We used WILMA as the main backend, covering the full 1 GHz bandwidth 
provided by HERA with 2 MHz of spectral resolution. We also used in parallel the versatile 
VESPA spectrometer to get some interesting lines within the 1 GHz range with higher spectral 
resolution (320 kHz), as we did for $J$=5--4 SiO (from all the 2D survey data
only the map of this line is showed in this paper). A full description 
of the 2D line survey will be published
elsewhere (Marcelino et al., in preparation).

\section{Results}
\label{sect_res}
The line survey has been presented elsewhere (Paper I). This
paper is devoted to the study of the emission of Silicon-bearing molecules: we
have detected SiO, $^{29}$SiO, $^{30}$SiO, SiO $\textit{v}$=1, SiS, 
$^{29}$SiS, and SiS $\textit{v}$=1.
For Si$^{18}$O and Si$^{17}$O we have evidences of their 
presence but all their lines
in our frequency range are blended with other molecules.
We report the detection of the transition J=4-3 of SiO $v$=1 for the first 
time in this source. While this transitions shows strong maser emission, rotational lines for 
$v$$\ge$2 in the covered frequencies lack of maser emission. In the following subsections
we describe the data and the results for each molecular species.

\subsection{SiO}
\label{sect_res_sio}

\begin{table*}
\begin{center}
\caption{Lines of SiO, SiO isotopologues, and SiO vibrationally excited\label{tab_siolines}}
\begin{tabular}{llllllll}
\hline
\hline 
Molecule & Observed & $T_{MB}$ & $\int T_{MB} dv$ & Trasition & Rest & E$_u$/k &
S$_{ij}$\\  & v$_{LSR}$ (km s$^{-1}$) & (K) & (K km s$^{-1}$) & J & Frequency (MHz) &
(K) &  \\
\hline
SiO & 11.0 & 17.5 & 394$\pm$18 & 2-1 & 86846.971 & 6.3 & 2.00 \\
 & 8.5 & 25.6 & 590$\pm$18 & 3-2 & 130268.665 & 12.5 & 3.00 \\
 & 8.5 & 44.6 & 962$\pm$29 & 4-3 & 173688.210 & 20.8 & 4.00 \\
 & 7.3 & 45.0 & 1485$\pm$22 & 5-4 & 217104.889 & 31.3 & 5.00 \\
 & 8.0 & 61.1 & 2219$\pm$43 & 6-5 & 260517.985 & 43.8 & 6.00 \\
$^{29}$SiO & 11.5 & 1.27 & ... & 2-1 & 85759.202 & 6.2 & 2.00 \\
 & 11.3 & 9.22 & 238$\pm$20 & 4-3 & 171512.814 & 20.6 & 4.00 \\
 & 15.5\tablefootmark{1} & 9.40 & ... & 5-4 & 214385.778 & 30.9 & 5.00 \\
 & 10.6\tablefootmark{2} & 12.9 & ... & 6-5 & 257255.249 & 43.2 & 6.00 \\
$^{30}$SiO & 14.9\tablefootmark{3} & 1.41 & ... & 2-1 & 84746.193 & 6.1 & 2.00 \\
 & 10.1 & 6.17 & 108$\pm$5 & 4-3 & 169486.929 & 20.3 & 4.00 \\
 & 10.6 & 6.36 & 140$\pm$4 & 5-4 & 211853.546 & 30.5 & 5.00 \\
 & 10.0 & 12.6\tablefootmark{4} & 153$\pm$14 & 6-5 & 254216.751 & 42.7 & 6.00 \\
Si$^{18}$O & 6.9\tablefootmark{5} & 0.26 & ... & 2-1 & 80704.907 & 5.8 & 2.00 \\
 & 8.6 & 4.30\tablefootmark{5}$^,$\tablefootmark{6} & ... & 4-3 & 161404.866 & 19.4 & 4.00 \\
 & 11.2 & 0.65\tablefootmark{2} & ... & 5-4 & 201751.443 & 29.0 & 5.00 \\
 & \tablefootmark{7} & ... & ... & 6-5 & 242094.928 & 40.7 & 6.00 \\
Si$^{17}$O & 20.4\tablefootmark{8} & 0.27 & ... & 2-1 & 83588.641 & 6.0 & 2.00 \\
 & 12.3 & 0.40 & ... & 4-3 & 167171.974 & 20.1 & 4.00 \\
 & 12.7 & 0.71\tablefootmark{9} & ... & 5-4 & 208959.990 & 30.1 & 5.00 \\
 & 9.2 & 0.55 & ... & 6-5 & 250744.688 & 42.1 & 6.00 \\
SiO $\textit{v}=1$ & 14.2 & 77.5 & 587$\pm$20 & 2-1 & 86243.430 & 6.2 & 2.00 \\
 & -3.9 & 113.5 & 750$\pm$19 & ... & ... & ... & ... \\
 & 14.0 & 21.5 & 123$\pm$14 & 4-3 & 172481.125 & 20.7 & 4.00 \\
 & -4.5 & 4.20 & ... & ... & ... & ... & ... \\
 & \tablefootmark{10} & $<$1.2 & ... & 5-4 & 215596.029 & 31.0 & 5.00 \\
 & \tablefootmark{2} & $<$1.2 & ... & 6-5 & 258707.350 & 43.5 & 6.00 \\
\hline 
\end{tabular}
\end{center}
\tablefoot{
Emission lines of SiO,
  SiO isotopologues, and SiO vibrationally excited present in the
  frequency range of the Orion KL survey. 
Column 1 indicates the species, Col. 2
gives the observed (centroid) radial velocities, Col. 3 gives the
peak line temperature, Col. 4 the integrated intensity, Col. 5 the line transition, Col. 6 the
calculated rest frequencies, Col. 7 the energy of the upper level, and
Col. 8 gives the line strength.\\
\tablefoottext{1}{Blended with $^{34}$SH$_2$.}
\tablefoottext{2}{Blended with CH$_3$OCOH.}
\tablefoottext{3}{Blended with CH$_3$OH.}
\tablefoottext{4}{Blended with CH$_3$CH$_2$CN b type.}
\tablefoottext{5}{Blended with CH$_3$CH$_2$CN in the plane torsion (Tercero et
  al., in preparation).}
\tablefoottext{6}{Blended with CH$_2$CHCN.}
\tablefoottext{7}{Blended with CH$_3$CH$_2$CN a type.}
\tablefoottext{8}{Blended with H$_{53}$$\beta$ and U line.}
\tablefoottext{9}{Blended with CH$_3$CH$_2$C$^{15}$N.}
\tablefoottext{10}{Blended with CH$_3$CH$_2$CN out of plane torsion (Tercero et
  al., in preparation).}
}
\end{table*}

\begin{figure*}
\includegraphics[angle=270,scale=.70]{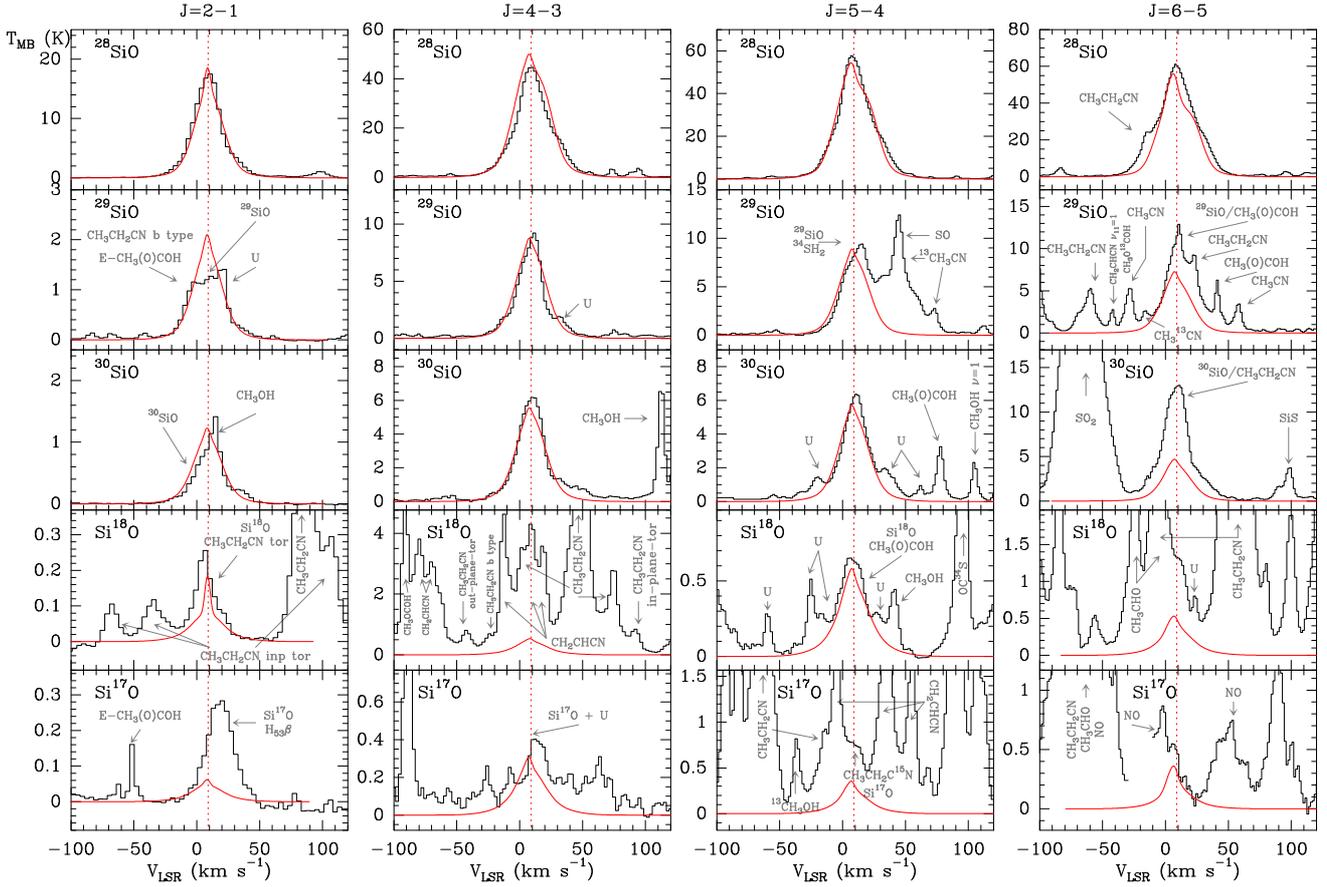}
\caption{Observed lines (histogram spectra) and model 
  (thin curves) of SiO and its isotopologues. The dashed line shows a radial velocity at
  9 km s$^{-1}$.}
\label{fig_sio}
\end{figure*}

\begin{figure*}
\includegraphics[angle=0,scale=1]{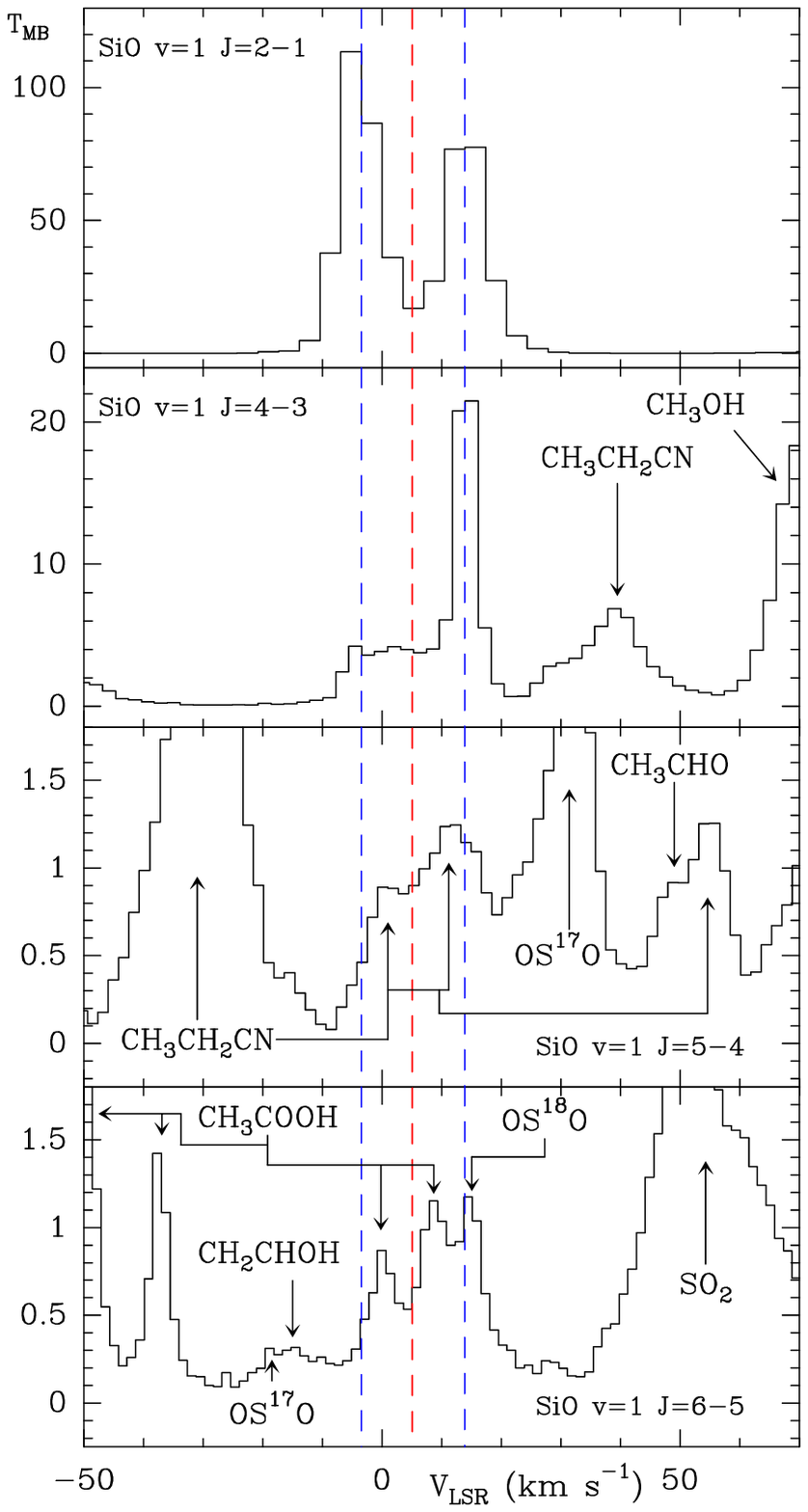}
\caption{Observed lines of vibrationally excited SiO $\textit{v}$ = 1
  showing maser emission in the transitions $J$ = 2-1 and $J$ = 4-3. The dashed 
  lines show radial velocities at
  at $-$3.4, $+$5.1 and $+$13.9 km s$^{-1}$.}
\label{fig_siov1}
\end{figure*}

Five rotational transitions of silicon monoxide (SiO) 
fall in the covered frequency range
of our line survey. For its
isotopologues ($^{29}$SiO, $^{30}$SiO, Si$^{18}$O, and Si$^{17}$O) and
SiO vibrationally excited four rotational transitions are within our
 frequency range.

The rotational constants used to derive the line parameters have been taken
from \citet{San03} (for SiO, $^{29}$SiO, $^{30}$SiO, Si$^{18}$O, 
SiO $\textit{v}$=1) 
and for Si$^{17}$O, the line parameters have been derived from the Dunham 
coefficients, Y$_{ij}$s, of $^{28}$SiO and the isotopic relations
Y$_{ij}$'=Y$_{ij}$*($\mu$(Si$^{17}$O)/$\mu$(SiO))$^{(i+2j)/2}$ (where
$\mu$ is the reduced mass of the isotopologue). 
The dipole moment ($\mu$=3.098D) was reported by
\citet{Ray70}. Line parameters
and observed intensities for all these lines are given in Table 
\ref{tab_siolines}. Figure \ref{fig_sio} shows the observed 
lines of SiO and its
isotopologues, together with the results from our best model 
(see Sect. \ref{sect_phy_sio}). 
Figure \ref{fig_siov1} shows the SiO $\textit{v}$=1 lines within
our line survey. The $J$ = 2-1 and $J$ = 4-3 lines show maser
emission whereas the $J$ = 5-4 and $J$ = 6-5 are blended with other 
abundant molecules in Orion and their possible contribution to the
observed features is rather weak and two orders of magnitude below
the observed emission in the $J$=2-1 and $J$=4-3 maser emission. 
Rotational lines from SiO $\textit{v}\ge$2 are below the
detection limit of this line survey. Only the $v$=2
$J$=2-1 seems to have a very week emission, T$_{MB}\simeq$0.06 K, 
at the velocity of the red component
of the maser.

The line profiles indicate the contribution to the emerging
intensities from different
velocity components: the extended and compact molecular ridge (difficult 
to separate), the hot core, the high velocity outflow, and the
low velocity plateau. The latter
appears as the most significant contribution to the line profile and
intensity. 

Table \ref{tab_sio_gau1}, available electronically only,
gives the parameters for the different
cloud components derived by fitting several Gaussian profiles to the observed
transitions of SiO and to selected lines of $^{29}$SiO and $^{30}$SiO.
The profiles are separated into two components: one wide component
which corresponds
to the low and high velocity outflows (the plateau), and a narrow one
consisting of a mix of the ridge components with the plateau.

The broad component from the plateau dominates the line profiles and hides
the contribution of the hot core at 5 km s$^{-1}$, which 
is only marginally detected as a blue
shoulder in the emission from the isotopologues. 
We interpret this behavior as due to
the opacity of the SiO lines in the high velocity gas which could
absorb the emission from the hot core (this effect was studied for HDO
by \citealt{Par01b}). However, the line profiles
of the $^{29}$SiO and $^{30}$SiO isotopologues, for which the high velocity
gas should be thinner, clearly need a component at the velocity of the
hot core (see also the discussion for SiS below) in order to reproduce their 
line profiles. 

\subsubsection{$J$=5--4 SiO map}
From the 2D survey data of Orion KL,
a map of the SiO $J$=5-4 intensity at different velocity ranges
is shown in Fig. \ref{fig_sio_map}.
The velocity structure of the SiO emission shows the contribution from
all the cloud components
quoted above. Note the spatial displacement of the emission peak
with velocity. Particularly interesting is the spatial distribution
of the red
and blue wings at the largest velocities (panels top left and bottom
right). The high velocity outflow
appears as an elliptical shell of gas around
IRc2.
Recently, \citet{Pla09} have obtained a map with an angular resolution
of $\simeq$0.5'' of the $v$=0 $J$=2-1 line of SiO. 
Their data for the extreme red
velocities, 29-52 kms$^{-1}$, indicates that the emission is shifted towards
the East from source $I$ by 5-10''. However, blue extreme velocities,
-20 to -40 km s$^{-1}$, are found several arcseconds W and NW of source $I$.
\citet{Wri96} obtain a similar result in their aperture synthesis ($\simeq$4''
resolution) $v$=0 $J$=2-1 SiO
velocity map of the extreme velocities 
(v$_{LSR}$$<$$-$11 km s$^{-1}$ and v$_{LSR}$$>$29 km s$^{-1}$).
These results are in very good agreement with our lower angular resolution map. 
\citet{Pla09} found that the bulk of the emission arises from a bipolar outflow 
covering velocities from
-13 to 16 km s$^{-1}$, driven by source $I$ and with an extent of 
$\simeq$6'' along the NE-SW 
direction. The large velocity outflow, or extreme velocities, 
seems to be a continuation of the 
low velocity outflow but less spatially structured. 
This ''EW –bipolarity'' of the SiO high velocity outflow was noted already 
by \citet{Olo81}.
We have obtained
angular source sizes between 16'' for the central velocities to 23'' for the
extreme velocities, assuming emission within the half flux level
and corrected for the size of the telescope beam at the observing frequency.
The distribution of the high velocity gas is very similar
to that found in the extended maser emission of H$_2$O at 183.3 GHz 
found by \citet{Cer90,Cer94}.

\subsection{SiS}
\label{sect_res_sis}

\begin{table*}
\begin{center}
\caption{Lines of SiS, $^{29}$SiS, and SiS $\textit{v}$=1 \label{tab_sislines}}
\begin{tabular}{llllllll}
\hline
\hline 
Molecule & Observed & $T_{MB}$ & $\int T_{MB} dv$ & Trasition & Rest & E$_u$/k &
S$_{ij}$\\ & v$_{LSR}$ (km s$^{-1}$) & (K) & (K km s$^{-1}$) & J & Frequency (MHz) &
(K) &  \\
\hline
SiS & 12.5 & 0.29 & 6.1$\pm$0.3 & 5-4 & 90771.558 & 13.1 & 5.00 \\
 & 16.7 & 0.53 & 11.5$\pm$0.5 & 6-5 & 108924.294 & 18.3 & 6.00 \\
 & 15.5 & 1.34 & 30$\pm$1 & 8-7 & 145227.045 & 31.4 & 8.00 \\
 & 16.5 & 2.14\tablefootmark{1} & 43$\pm$1 & 9-8 & 163376.773 & 39.2 & 9.00 \\
 & 16.2 & 2.42 & ... & 11-10 & 199672.219 & 57.5 & 11.00 \\
 & 11.1 & 3.10 & ... & 12-11 & 217817.651 & 68.0 & 12.00 \\
 & 8.9\tablefootmark{2} & 20.0 & ... & 13-12 & 235961.363 & 79.3 & 13.00 \\
 & 17.7 & 4.45 & ... & 14-13 & 254103.213 & 91.5 & 14.00 \\
 & 15.2 & 4.25 & ... & 15-14 & 272243.058 & 104.5 & 15.00 \\
$^{29}$SiS & 16.6 & 0.019 & ... & 5-4 & 89103.781 & 12.8 & 5.00 \\
 & \tablefootmark{3} & ... & ... & 6-5 & 106923.019 & 18.0 & 6.00 \\
 & 16.1 & 0.087 & ... & 8-7 & 142558.873 & 30.8 & 8.00 \\
 & \tablefootmark{4} & ... & ... & 9-8 & 160375.213 & 38.5 & 9.00 \\
 & \tablefootmark{5} & ... & ... & 12-11 & 213816.228 & 66.7 & 12.00 \\
 & 13.1 & 0.25 & ... & 13-12 & 231626.772 & 77.8 & 13.00 \\
 & \tablefootmark{6} & ... & ... & 14-13 & 249435.521 & 89.8 & 14.00 \\
 & 12.9 & 0.24 & ... & 15-14 & 267242.338 & 102.6 & 15.00 \\
SiS $\textit{v}$=1 & noise level & ... & ... & 5-4 & 90329.891 & 13.0 & 5.00 \\
 & \tablefootmark{2} & ... & ... & 6-5 & 108394.292 & 18.2 & 6.00 \\
 & 12.4 & 0.12 & ... & 8-7 & 144520.370 & 31.2 & 8.00 \\
 & 13.3 & 0.46\tablefootmark{7} & ... & 9-8 & 162581.760 & 39.0 & 9.00 \\
 & 11.9 & 0.17 & ... & 11-10 & 198700.526 & 57.2 & 11.00 \\
 & \tablefootmark{8} & ... & ... & 12-11 & 216757.615 & 67.6 & 12.00 \\
 & 14.8 & 0.31 & ... & 13-12 & 234812.983 & 78.9 & 13.00 \\
 & 13.8 & 0.36 & ... & 14-13 & 252866.487 & 91.0 & 14.00 \\
 & \tablefootmark{1} & ... & ... & 15-14 & 270917.984 & 104.0 & 15.00 \\
\hline
\end{tabular}
\end{center}
\tablefoot{
Emission lines of SiS, $^{29}$SiS, and SiS $\textit{v}$=1 present in the
frequency range of the Orion KL survey. Column 1 gives the species, Col. 2
the observed (centroid) radial velocities, Col. 3 gives the
peak line temperature, Col. 4 the integrated intensity, Col. 5 the line transition, Col. 6 the
calculated rest frequencies, Col. 7 the energy of the upper level, and
Col. 8 gives the line strength.\\
\tablefoottext{1}{Blended with CH$_3$OCOH.}
\tablefoottext{2}{Blended with $^{13}$CH$_3$OH.}
\tablefoottext{3}{Blended with U line.}
\tablefoottext{4}{Blended with (CH$_3$)$_2$CO.}
\tablefoottext{5}{Blended with $^{34}$SO$_2$.}
\tablefoottext{6}{Blended with CH$_3$OH.}
\tablefoottext{7}{Blended with $^{33}$SO$_2$.}
\tablefoottext{8}{Blended with CH$_3$CH$_2$CN b type.}
}
\end{table*}

\begin{figure*}
\includegraphics[angle=0,scale=.85]{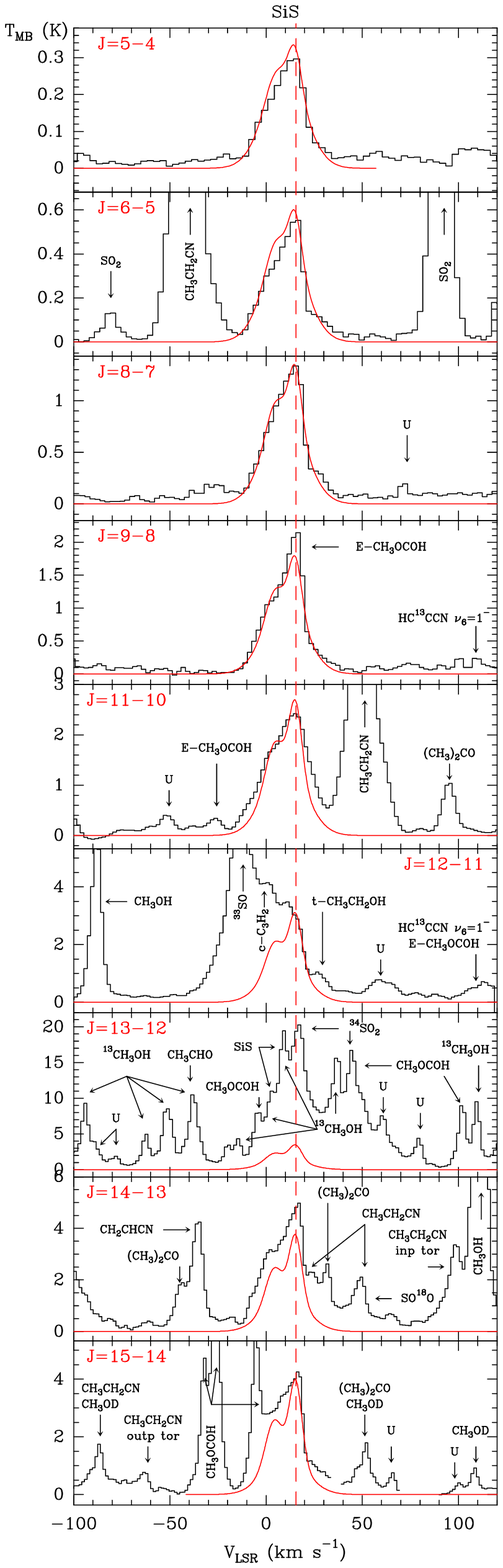}
\caption{Observed lines (histogram spectra) and model 
  (thin curves) of SiS. The dashed line shows a radial velocity at
  15.5 km s$^{-1}$.}
\label{fig_sis}
\end{figure*}

Nine transitions of silicon monosulfide $^{28}$Si$^{32}$S $\textit{v}$=0,1
and of its rare isotopologues
are present in the covered frequency range.
Spectroscopic constants are from \citet{San03}.
The dipole moment ($\mu$=1.730D) has been taken
from \citet{Hoe69}.
Line frequencies and observational
line parameters are given in Table
\ref{tab_sislines}. The observed line profiles and the results from
our best model (see Sect. \ref{sect_phy_sis}) are shown in Fig.
\ref{fig_sis}, for the main isotopologue, and in Fig. \ref{fig_29sis_sisv1}
(for $^{29}$SiS and SiS $\textit{v}$=1).
The lines from the other isotopologues are blended with strong lines
arising from other molecular species. Only one line ($J$=9-8 transition) is free of
blending for Si$^{34}$S. The line intensities arising from the
isotopologues $^{30}$SiS and Si$^{33}$S are below the confusion limit
of our line survey.

The line profiles of the most abundant isotopologue display three 
components: a wide component that corresponds to the 
plateau, the hot core (clearly seen in the profile of the
$J$ = 14-13 transition), and
a narrower one at v$_{LSR}$ $\simeq$ 15.5
km s$^{-1}$. No emission has been observed from the ridge component.
Line parameters are given in Table \ref{tab_sis_gau1},
available electronically only (in the wide 
component emission from the plateau and the hot core are merged).

The velocity of the emission peak for the main isotopologue is at 
v$_{LSR}$ $\simeq$ 15.5 km s$^{-1}$ which coincides with the LSR velocity 
of the red component of the SiO $\textit{v}$=1
maser emission (which could be a fortuitous agreement, see
Sect. \ref{sect_res_fea}). 
Previous works discuss the presence of SiS in Orion KL
and its association with the SiO $\textit{v}$=1 maser \citep{Dic81,
Sut85, Ziu88, Ziu91, Sch97}.

Although confusion is large when considering the weak lines of
$^{29}$SiS and SiS $\textit{v}$=1, we found that their emission peaks
at v$_{LSR}$ $\simeq$ 13.5 km s$^{-1}$ (see Fig. \ref{fig_29sis_sisv1}),
a mixture of all cloud
components, but dominated by the 15.5 km s$^{-1}$ feature.

\subsubsection{The feature at 15.5 km s$^{-1}$}
\label{sect_res_fea}

\begin{figure*}
\includegraphics[angle=0,scale=.55]{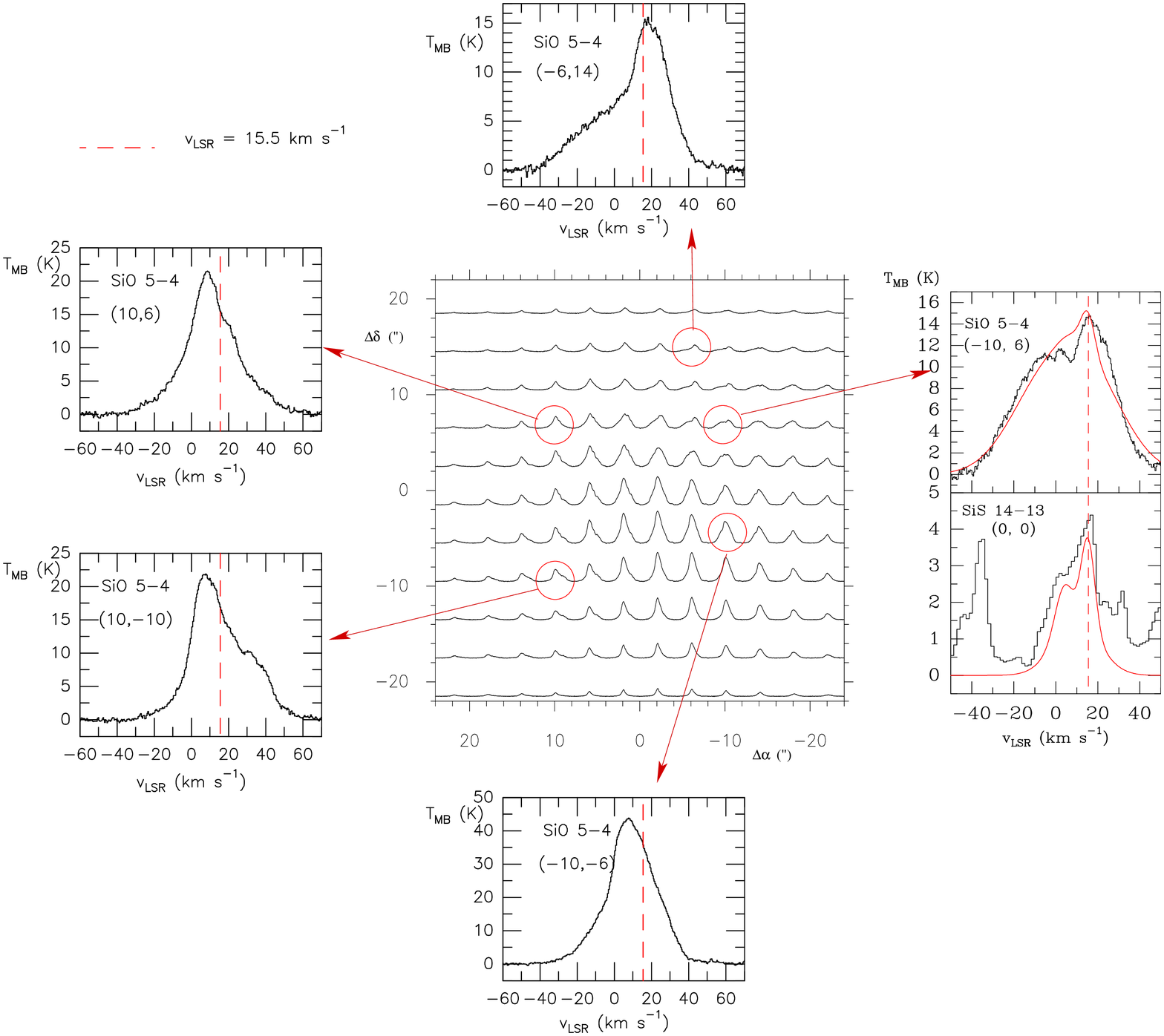}
\caption{{\it{Central panel}} shows a
20'' x 20'' map, centered at IRc2, of the $J$=5-4 line of SiO;
the {\it{right panel}} shows a comparison between the $J$=5-4 line 
at an offset (-10'',6''), and the SiS J=14-13 
line towards IRc2, and the resulting model for each line (thin curves). 
The {\it{rest of the panels}} show the 
SiO J=5-4 line at different positions.}
\label{fig_sio_sis}
\end{figure*}

At the position of the line survey (IRc2) only SiS
(\textit{v}=0, 1) lines and one component
of the SiO maser emission (\textit{v}=1) show an intensity peak
at 15.5 km s$^{-1}$.
\citet{Wri90} measured the absolute 
position of the SiO masers to an accuracy of 0''.15 
coinciding with the position 
of radio source $I$ (very close to IRc2).
In order
to check the origin of this feature we have observed the line
profiles of several molecules around IRc2. We have found
that the line profiles of SiO show important changes with position.
Figure \ref{fig_sio_sis} central panel displays a 20'' x 20'' map centered 
on IRc2 of the $J$=5-4 line of SiO;
the small panels around the map show the line profile of the $J$=5-4 transition
of SiO at selected positions and show strong differences across the cloud. 
The intensity peak of SiO lines is at 
15.5 km s$^{-1}$ at the positions $\Delta\alpha$=-10'',
$\Delta\delta$=+6'' and  $\Delta\alpha$=-6'',
$\Delta\delta$=+14''. 
A comparison of the $J$=5-4 SiO line at (-10'', 6'') with the  
emission of the $J$=14-13 SiS line at (0, 0) is shown in the right panel.

In our map, this extended feature  
is seen around the position 
$\Delta\alpha$=-7'', $\Delta\delta$=+7'' (near the BN object) with
a radius of $\simeq$5 arcseconds.  
Figure \ref{fig_m15_15} shows emission lines from different molecules
at a position offset (-15'', 15'') from IRc2. 
Only molecules with a strong emission from the plateau component (SO,
$^{34}$SO, SO$_2$, and SiO) show the 15.5 km s-1 component at this 
position. Hence, this feature is not a particularity 
of SiS but arises probably from the interaction of the outflow with
the ambient cloud. Consequently, lines toward Orion KL showing a 
strong intensity and eventually high opacity in the plateau could 
hide interesting details of other components 
(hot core, feature at 15.5 km s-1), making line interpretation a 
very difficult task.

\section{Physical parameters of the clouds}
\label{sect_phy}

\begin{table*}
\begin{center}
\caption{Physical parameters of Orion KL components\label{tab_prop}}
\begin{tabular}{lllllll}
\hline 
\hline 
 Parameter & Extended ridge & Compact ridge & Plateau & High velocity plateau & Hot core & 15.5 km s$^{-1}$ component\\
\hline
Source diameter ('') & 120 & 15 & 30 & 30 & 10 & 10\\
Offset (IRc2) ('') & 0 & 7 & 0 & 0 & 2 & 5\\
n(H$_2$) (cm$^{-3}$) & 1.0$\times$10$^{5}$ &
1.0$\times$10$^{6}$ & 1.0$\times$10$^{6}$ & 1.0$\times$10$^{6}$ & 
5.0$\times$10$^{7}$ & 5.0$\times$10$^{6}$\\
T$_K$ (K) & 60 & 110 & 125 & 125 & 225 & 200\\
v$_{half\;\;power\;\;intensity}$ (km s$^{-1}$) & 4 & 4 & 25 & 50 & 10 & 7.5\\
v$_{LSR}$ (km s$^{-1}$) & 9 & 7.5 & 9 & 9 & 5.5 & 15.5\\
\hline 
\end{tabular}
\end{center}
\tablefoot{
Obtained physical parameters for Orion KL.
}
\end{table*}


Column densities for all detected species have been calculated using
an excitation and radiative transfer code developed by J. Cernicharo
(Cernicharo 2010, in preparation). Depending on the selected molecule
or physical conditions, we assume the large velocity gradient (LVG; 
\citealp{Sob58}; \citealp{Sob60}) or local thermodynamic equilibrium
(LTE) approximations.
Table \ref{tab_prop} resume the physical parameters we have obtained
for each spectral cloud component 
from the modeling of SiO and SiS (note that for the
new feature at 15.5 km s$^{-1}$ we have derived the following
parameters from the modeling of the SiS lines: 
T$_K$=200 K, n(H$_2$)=5$\times$10$^6$ cm$^{-3}$, and
v$_{half\;\;power\;\;intensity}$=7.5 km s$^{-1}$).
We assume uniform physical conditions: kinetic temperature, density, radial
velocity, and line width.
We adopted these values from the data 
analysis (Gaussian fits and an attempt to simulate the line widths and
intensities with LTE and LVG codes) as representative parameters
for the different species.
Our modeling also takes into account the size
of each component and its offset position with respect to IRc2. 
Corrections for beam dilution are applied for each line depending 
on their frequency. 

The only free parameter is, therefore, the column density 
of the corresponding observed species. Taking into account the reduced
size of most cloud components the contribution from the error beam is
negligible except for the extended ridge which has a small
contribution for all observed lines.

In addition to line opacity effects, we discussed other sources 
of uncertainty in Paper I. 

%



\subsection{SiO}
\label{sect_phy_sio}

\begin{table*}
\begin{center}
\caption{Column densities - SiO \label{tab_cdsio}}
\begin{tabular}{llllll}
\hline 
\hline 
 Species & Extended ridge & Compact ridge & Plateau & High velocity plateau & Hot core \\
 & N $\times$10$^{15}$ (cm$^{-2}$) & N $\times$10$^{15}$ (cm$^{-2}$) &
N $\times$10$^{15}$ (cm$^{-2}$) & N $\times$10$^{15}$ (cm$^{-2}$) 
& N $\times$10$^{15}$ (cm$^{-2}$)\\
\hline
SiO from $^{29}$SiO & 0.04$\pm$0.01 & 0.20$\pm$0.05 &
5.0$\pm$1.0 & 1.4$\pm$0.4 & 1.0$\pm$0.3 \\
SiO from $^{30}$SiO & 0.03$\pm$0.007 & 0.15$\pm$0.04 &
4.5$\pm$1.0 & 1.5$\pm$0.4 & 1.2$\pm$0.3\\
SiO (Average) & 0.035$\pm$0.009 & 0.17$\pm$0.05 &
4.7$\pm$1.0 & 1.4$\pm$0.4 & 1.1$\pm$0.3\\
$^{29}$SiO & 0.0020$\pm$0.0005 & 0.010$\pm$0.003 &
0.25$\pm$0.06 & 0.07$\pm$0.02 & 0.05$\pm$0.02 \\
$^{30}$SiO & 0.0010$\pm$0.0003 & 0.005$\pm$0.002 &
0.15$\pm$0.04 & 0.05$\pm$0.02 & 0.04$\pm$0.01 \\
Si$^{18}$O & $\lesssim$0.0005
& $\lesssim$0.0005 & $\lesssim$0.01 & $\lesssim$0.01 & $\lesssim$0.01 \\
Si$^{17}$O & $\lesssim$0.0001
& $\lesssim$0.0005 & $\lesssim$0.005 & $\lesssim$0.005 & $\lesssim$0.005 \\
\hline 
\end{tabular}
\end{center}
\tablefoot{
Column densities of the different SiO isotopologues calculated with LVG and LTE codes.
}
\end{table*}

LTE conditions have been assumed for the hot core, while
LVG calculations have been performed for the extended ridge, compact ridge,
plateau, and high velocity plateau, with collisional cross sections of 
SiO-p-H$_2$ taken from 
\citet{Day06} (see Appendix) for temperatures between 10 K to 
300 K including levels 
up to $J$ = 20 (E$_{up}$=437 K).

SiO lines appear to be optically thick and therefore the derived 
SiO column density could be significantly underestimated. 
The lines of the isotopologues are, however, 
mostly optically thin so we can estimate the SiO column density 
assuming standard isotopic abundance ratios 
($^{28}$Si/$^{29}$Si$\simeq$20 and
$^{28}$Si/$^{30}$Si$\simeq$30; \citealt{And89}). 
The results obtained are shown in Table \ref{tab_cdsio}. 
Due to the weakness of the less abundant isotopologues 
lines (Si$^{18}$O and Si$^{17}$O) and large line overlap problems, 
we can only obtain upper limits for their column density.
We have estimated the uncertainty between 20-30 \% for the 
$^{29}$SiO and $^{30}$SiO results.

The cloud component with the largest column density corresponds to the  
plateau with $N$(SiO)$_{plateau}$ = (4.7$\pm$1.0)$\times$10$^{15}$
cm$^{-2}$; while the total column density of SiO is
$N$(SiO)=(7.4$\pm$2.0)$\times$10$^{15}$ cm$^{-2}$, both results in
agreement with those obtained from the line survey of Orion
using the Odin satellite \citep{Per07}.
Our study provides a column density between six times to one order
of magnitude larger than that obtained in the surveys 
at high frequency by
\citet{Sch01} and \citet{Com05}. Larger column densities
have been reported by \citet{Ziu87} and \citet{Wri96}, while
\citet{Joh84} and \citet{Sut95} obtained a beam average column density in 
good agreement with our results. These differences are mostly due to
the different
assumptions on the physical conditions and cloud structure made
in the interpretation of the observations.

We derive a column density $^{29}$SiO/$^{30}$SiO ratio of 2, 2, 1.7,
1.4, 1.25 for the extended ridge, compact ridge, plateau, high
velocity plateau, and hot core, respectively, in agreement with the
standard solar value of $^{29}$Si/$^{30}$Si $\simeq$ 1.5 \citep{And89}. 
We can estimate a lower limit for the isotopic ratio $^{16}$O/$^{18}$O
and an estimation of $^{18}$O/$^{17}$O by means of the total 
column density ratios
$N$(Si$^{16}$O)/$N$(Si$^{18}$O) $\ga$ 239 and
$N$(Si$^{18}$O)/$N$(Si$^{17}$O) $\simeq$ 2 with large uncertainty due
to the severe blending of their lines with other species. These values
are two times lower than those of the solar system. In Paper I we derived
$N$($^{16}$OCS)/$N$($^{18}$OCS) = 250$\pm$135, while \citet{Per07} obtained
a $^{18}$O/$^{17}$O ratio of 3.6 from C$^{18}$O/C$^{17}$O.  

We have also modeled the 5-4 SiO
line in the offset position (-10'',6'') from IRc2
(see Fig. \ref{fig_sio_sis}) in order to provide a
column density for the 15.5 km s$^{-1}$ feature. Using the same column 
densities obtained above for the different components, we have added 
the feature at 15.5 km s$^{-1}$ with a column density of 
(1.0$\pm$0.3)$\times$10$^{15}$ cm$^{-2}$.  

\subsubsection{SiO maser lines}
\label{sect_phy_mas}

SiO maser emission in Orion was discovered by \citet{Sny74}
through its $v$=1 $J$=2-1 transition. \citet{Pla90}
suggested that the emission arises from an expanding rotating disk
around IRc2. The emission from this line, studied by
several authors, has been found to be confined to a region of 20-100 AU  
(3-10$\times$10$^{14}$ cm) around the radiocontinuum source $I$ \citep{Chu87}
and to trace a protostellar wind and/or outflow expanding with
a velocity $\simeq$ 20 kms$^{-1}$ \citep{Pla90, Men95,
Gre98, Doe99, Pla03, Doe04, Gre04, Mat07}.
Source $I$ has not an infrared counterpart in the range
4-22 $\mu$m \citep{Gre04}, suggesting that it is surrounded by 
a dusty disk producing severe extinction. Source $I$ is placed between
Orion IRc2 and the hot core and excites the masers of SiO and H$_2$O
\citep{Gen89, Pla95}. 
Several other lines of SiO in $v$=1 and 2 have been
searched towards this source an some of them have been detected
($v$=1 $J$=1-0, $J$=2-1, $J$=3-2; see references above). The $v$=1 $J$=4-3 line
was searched without success by \citet{Sch82}. In our line survey
we have covered the $J$=2-1, 4-3, 5-4, and 6-5 lines of all vibrational
states of SiO. As shown in Fig. \ref{fig_siov1}, 
only the $v$=1 $J$=2-1 and 4-3 transitions 
have been unambiguously detected (the $J$=4-3 for the first time).
For the rotational lines of the $v$=2 level the range of velocities 
covered by the SiO $v$=1 $J$=2-1
maser is always contaminated by other lines. 
The $v$=2 $J$=2-1 line shows a weak feature
coincident with the red emission at 15 km s$^{-1}$ but it arises from the
4$_{2,3}$-3$_{1,2}$ line of E-CH$_3$OCOH as confirmed 
from the observed intensities
of other lines of this species. The $v$=2 $J$=4-3 line shows a strong
feature at the velocity of the blue component of the maser but it arises from
(CH$_3$)$_2$CO. The other lines from $v$=2, 3, and 4 are always blended with 
other lines. In our survey we have covered the same rotational and vibrational
quantum numbers of the SiO isotopologues. 
None of them shows any maser effect
within the sensitivity of our observations.

The observation of several ro-vibrational
lines of SiO can be used to constraint the physical properties of
the gas at spatial scales from a few tens to a few hundred AU from the
source. \citet{God09} have modeled the emission of several isotopologues
of SiO. They conclude that while the $v$=0 $J$=1-0 maser emission
arises from a region with radius $>$100 AU and densities $<$10$^{7}$ cm$^{-3}$,
the $v$=1,2 $J$=1-0 and $J$=2-1 are produced in a region of $\simeq$100 AU
around source $I$. Our observed intensity ratio $v$=1 $J$=2-1/$J$=4-3 is
$\simeq$5-10 depending on the velocity component. We obtain such a behavior
for column densities of $\simeq$10$^{19}$ cm$^{-2}$ and densities
$\simeq$ 10$^8$ cm$^{-3}$. For higher densities the ratio increases and the
predicted intensities of the $J$=2-1 and $J$=4-3 lines, although large,
are not compatible with the inferred brightness temperatures which could
be well above 10$^5$ K. We have run several models with different physical
parameters. Our results suggest that the gas responsible for maser emission
could have densities up to 10$^8$ cm$^{-3}$, sizes of 100 AU, and gas
temperatures around 600-800 K, a factor two below those of
\citet{God09}. With these parameters for the emitting region the
abundance of SiO should be as high as 10$^{-4}$.

\citet{Gon97} have shown that the masers
of SiO and their isotopologues are affected by line overlap in the
infrared between themselves. Such overlaps seem to have little effects
on the emission of SiO, $^{29}$SiO, and $^{30}$SiO in $v$$\neq$0 for
which maser emission has been found in evolved stars \citep{Cer91b, Gon96}.

\subsection{SiS}
\label{sect_phy_sis}

\begin{table*}
\begin{center}
\caption{Column densities - SiS \label{tab_cdsis}}
\begin{tabular}{llll}
\hline 
\hline 
 Species & 15.5 km s$^{-1}$ Component & Plateau & Hot core \\
 & N $\times$10$^{14}$ (cm$^{-2}$) & N $\times$10$^{14}$ (cm$^{-2}$) &
N $\times$10$^{14}$ (cm$^{-2}$)\\
\hline
SiS & 7.0$\pm$1.7 & 3.5$\pm$0.8 & 3.0$\pm$0.7 \\
$^{29}$SiS & 0.50$\pm$0.25 & 0.10$\pm$0.05 & 0.10$\pm$0.05 \\
SiS $v$=1 & 0.80$\pm$0.40 & 0.20$\pm$0.10 & 0.20$\pm$0.10 \\
\hline 
\end{tabular}
\end{center}
\tablefoot{
Column densities of different SiS species calculated with 
LVG and LTE codes.
}
\end{table*}

In the same way as for SiO, LTE conditions have been assumed 
for the hot core, while
LVG calculations have been performed for the 15.5 km s$^{-1}$ feature and
the plateau, with collisional cross sections SiS-o-H$_2$ for 41 levels 
and 5$<$T$_k$$<$300 K taken from \citet{Klo08} (see Appendix).
The fits are shown in Figs. \ref{fig_sis} and \ref{fig_29sis_sisv1}. 
The column density results are shown in Table
\ref{tab_cdsis}. We have estimated the uncertainty to be about 20-30 \%
for the results of SiS and 50 \% for $^{29}$SiS and SiS $v$=1. 
Our largest value of the column density for SiS corresponds to the
feature at 15.5 km s$^{-1}$ obtaining
$N$(SiS)$_{15.5\;\;kms^{-1}\;\;feature}$
 = (7.0$\pm$1.7) $\times$ 10$^{14}$ cm$^{-2}$;
the total column density for SiS is
$N$(SiS)$_{total}$ = (1.35$\pm$0.40) $\times$ 10$^{15}$
cm$^{-2}$. A similar value was obtained by \citet{Ziu88} and
\citet{Ziu91} whereas \citet{Dic81} found $N$(SiS) = (1-2) $\times$ 10$^{13}$
cm$^{-2}$.

Assuming that SiS emission is optically thin, as indicated by our 
calculations, we derive an isotopic abundance of
$^{28}$Si/$^{29}$Si$\simeq$14, 35, and 30 for the 15.5 feature, the
plateau, and the hot core, respectively
$\textit{i. e.}$ close to the solar system value. We provide an
average value of $^{28}$Si/$^{29}$Si = 26$\pm$10.

The $N$(SiO)/$N$(SiS) column density ratio observed in the plateau 
is $\simeq$13, in good agreement with \citet{Dic81} and four times
lower than the cosmic O/S ratio of 48 \citep{And89}. \citet{Ziu91}
found a $N$(SiO)/$N$(SiS) ratio $\simeq$ 40-80.
We can also derive this ratio by means of
$N$($^{29}$SiO)/$N$($^{29}$SiS), obtaining a value $\simeq$ 25 for the plateau.
In order to compare column density ratios, we have to
assume that the region of the line formation is the same for each molecule
and the excitation temperature is similar for both species, for that
reason we only provide this ratio for the plateau component.
In the previous work of this line survey (Paper I), 
we derived O/S ratios using different species/families 
of molecules: 
$N$(HCO$^+$)/$N$(HCS$^+$) $\simeq$ 13, $N$(H$_2$CO)/$N$(H$_2$CS)
$\simeq$ 12, and $N$(CO)/$N$(CS) $\simeq$ 370; in addition, \citet{Per07} found
$N$(H$_2$O)/$N$(H$_2$S) $\simeq$ 20 and $N$(H$_2$CO)/$N$(H$_2$CS)
$\simeq$ 15. All the obtained values, but $N$(CO)/$N$(CS) $\simeq$ 370,
show a similar O/S ratio, indicating that the different formation
paths of different molecules maintain a constant ratio O/S in the same
particular region of the cloud.

>From the column density obtained for SiS in the ground and the
vibrationally excited states, it is easy to estimate a vibrational
temperature by means of:

\begin{equation}
\frac{exp \left( \begin{array}{c} - 
\frac{E_{v=x}}{T_{vib}}\end{array} 
\right)}{f_{\nu}} = \frac{N
  (SiS\;\;\textit{v}=x)}{N (SiS)}
\end{equation} 
where E$_{\nu_x}$ is the excitation energy of the vibrational state 
(E$_{\textit{v}={1}}$ = 1077 K, $T_{vib}$ is the vibrational 
temperature, $f_{\nu}$ is the vibrational partition function, 
N(SiS $\textit{v}=x$) is the column
density of the vibrational state and N(SiS) is the column density of
SiS in the ground state. The vibrational partition function can be 
approximated by

\begin{eqnarray}
f_{\nu} = 1 + exp \left( \begin{array}{c} -
  \frac{E_{\nu_{3}}}{T_{vib}} \end{array} \right) + \nonumber \\
 + 2 exp \left( \begin{array}{c} -
  \frac{E_{\nu_{2}}}{T_{vib}} \end{array} \right) + 
exp \left( \begin{array}{c} -
  \frac{E_{\nu_{1}}}{T_{vib}} \end{array} \right)  
\end{eqnarray}\\
which, for low T$_{vib}$ leads to f$_{\nu}$ $\simeq$ 1.

We obtain T$_{vib}$ =
500$\pm$200 K in the 15.5 km s$^{-1}$ component. This value is
higher than the kinetic temperature we have assumed for that 
component (200 K). This result could indicate an inner and 
hotter emitting region for vibrationally excited SiS,
suggesting that the excitation temperature varies across
the feature at 15.5 km s$^{-1}$.
In Paper I we calculated the vibrational temperature in
the hot core component for OCS $\nu_2$ = 1 and $\nu_3$ = 1 
obtaining $\simeq$210 K and an upper limit of 300 K for CS $v$ = 1.
As we have indicated in our previous paper, all these results
could point to radiative pumping effects in the populations of 
the vibrationally excited states of these molecules (taking into
account the magnitude of the collisional rates of these species).

Higher angular resolution observations are necessary
to resolve any possible excitation gradient and temperature
profile in the feature at 15.5 km s$^{-1}$ and in the hot core
component.

\subsection{Other Silicon-bearing Molecules}
\label{sect_phy_oth}

\begin{table*}
\begin{center}
\caption{Column density upper limits\label{tab_nondetected}}
\begin{tabular}{llll}
\hline 
\hline 
 Molecule
 & Column density & Dipole & References\\
 & $\la$ N $\times$10$^{14}$ (cm$^{-2}$) & 
moment (D) & spectroscopic constants\\
\hline
 SiC              & 1.3 & 1.600$^1$ & (2)\\         
 SiC$_2$          & 0.35 & 2.393$^3$ & (4)\\
 c-SiC$_3$        & 0.13 & 4.200$^5$ & (6)\\
 SiC$_4$          & 0.04 & 6.420$^7$ & (8)\\
 SiN              & 0.61 & 2.560$^9$ & (10)\\
 SiCN             & 0.31 & 2.900$^{11}$ & (12)\\
 SiNC             & 0.31 & 2.000$^{11}$ & (12)\\
 ob-SiC$_3$       & 0.16 & 2.200$^{5}$ & (13)\\
 l-SiC3           & 0.04 & 4.800$^{14}$ & (14)\\
 Si$_3$           & 4 & 0.350$^{15}$ & (16)\\
 SiCCO            & 0.40 & 1.937$^{17}$ & (18)\\
 SiCCS            & 0.65 & 1.488$^{19}$ & (19)\\
 o-SiH$_2$        & 13 & 0.075$^{20}$ & (21)\\
 o-H$_2$CSi       & 1.7 & 0.300$^{22}$ & (22)\\
 p-H$_2$CSi       & 1.4 & 0.300$^{22}$ & (22)\\
 mb-Si$_2$H$_2$   & 0.34 & $\mu_a$=0.962/$\mu_b$=0.039$^{23}$ & (24)\\
 o-db-Si$_2$H$_2$ & 2.5 & $\mu_c$=0.480$^{25}$ & (25)\\
\hline 
\end{tabular}
\end{center}
\tablefoot{
Upper limits for the column density of non-detected
Silicon-bearing molecules in Orion KL.\\
(1) \citet{Lan90}; (2) \citet{Cer89} and \citet{Bog90};
(3) \citet{Sue89}; (4) \citet{Got89} and
\citet{Sue89}; (5) \citet{Alb90}; (6) \citet{App99b}; 
(7) \citet{Gor00}; (8) \citet{Ohi89} and \citet{Gor00};
(9) \citet{Ker05}; (10) \citet{Sai83} and \citet{Biz06};
(11) \citet{App00}; (12) \citet{App00} and \citet{McC01};
(13) \citet{McC99}; (14) \citet{McC00}; (15) \citet{Vas97};
(16) \citet{McC03a}; (17) \citet{Bot05}; (18) \citet{San05};
(19) \citet{Bot02}; (20) \citet{Gab93}; (21) \citet{Hir99};
(22) \citet{Izu96}; (23) \citet{Gre92};
(24) \citet{Cor92} and \citet{McC03b}; (25) \citet{Bog94}.
}
\end{table*}

We provide upper limits for the column density of several
Silicon molecules not detected in our line survey. We have assumed
the four typical spectroscopic components of Orion KL (hot
core, extended ridge, compact ridge, and plateau) plus the 15.5
feature, and a LTE approximation for all these molecules.
Table \ref{tab_nondetected} shows the results obtained, the dipole
moment of each species, and references for the spectroscopic constants.

{\it{SiC}}.- \citet{Mil80}
predicted SiC to be an abundant form of Silicon in dense
clouds on basis of gas phase chemistry models. This molecule has been 
detected by \citet{Cer89} towards the envelope of the red giant
star IRC+101216, but not in Orion (see
\citealp{Sch97} for the attempts of detection in this source).
We have obtained an upper limit for its total column density 
of 1.3$\times$10$^{14}$ cm$^{-2}$ providing  
an abundance ratio of $N$(SiO)/$N$(SiC)$\ga$57.

{\it{SiC$_2$}}.- We have not detected silicon dicarbide in our line
survey. \citet{Tur91} reported the detection of this
molecule in the hot core of Orion KL. However,
this result is quite uncertain due
to the few observed transitions (three) and their weakness. We have
searched for other lines of SiC$_2$ in our line survey and we
conclude that the intensities of the lines from
this molecule are below the confusion limit
as many of these lines are blended or missing.
We obtain an upper limit to its column density of
3.5$\times$10$^{13}$ cm$^{-2}$
and an abundance ratio $N$(SiO)/$N$(SiC$_2$)$\ga$211.
This molecule and its isotopologues have also been
detected towards IRC+101216 by \citet{Tha84} (SiC$_2$), \citet{Cer86}
($^{29}$SiC$_2$ and $^{30}$SiC$_2$) and 
\citet{Cer91a} (Si$^{13}$CC).

{\it{c-SiC$_3$}}.- \citet{App99a} detected rhomboidal SiC$_3$ in the expanding
envelope of the evolved carbon star IRC+10216.
For this molecule we
have calculated an upper limit to its column density in Orion KL of
1.3$\times$10$^{13}$ cm$^{-2}$, and a ratio $N$(SiO)/$N$(SiC$_3$)$\ga$570.

{\it{SiC$_4$}}.- SiC$_4$ was first detected in the space by
\citet{Ohi89} in the envelope of IRC+10216. 
For this molecule we
have calculated an upper limit to its column density in Orion KL of
3.5$\times$10$^{12}$ cm$^{-2}$, and a ratio $N$(SiO)/$N$(SiC$_4$)$\ga$2114.

{\it{SiN}}.- Silicon nitride has been detected in IRC+10216 by \citet{Tur92}
and in the galactic center cloud SgrB2(M) by \citet{Sch03}. We have
not found SiN above the confusion limit in Orion; note, however,
that some U-lines in our survey could be assigned to SiN (see
Fig. \ref{fig_sin}, panels 1, 6 and 9) but we
consider that the evidences for its presence in Orion are not strong
enough within the coverage of our survey. We provide an upper
limit of 6.1$\times$10$^{13}$ cm$^{-2}$, and $N$(SiO)/$N$(SiN)$\ga$121.

{\it{SiCN}}.- Cyanosilylidyne was identified in spectra recorded 
toward IRC+10216 by \citet{Gue00}. The calculated upper limit to its
column density in Orion KL is 3.1$\times$10$^{13}$ cm$^{-2}$, deriving
the ratio $N$(SiO)/$N$(SiCN)$\ga$240.

{\it{SiNC}}.- The isocyanosilylidyne isomer has a thermodynamic stability 
very similar to that of SiCN, but a slightly smaller dipole moment. 
It was detected toward IRC+10216 by \citet{Gue04}.
We provide an upper limit of 3.1$\times$10$^{13}$ cm$^{-2}$ to its 
column density, and a
ratio $N$(SiO)/$N$(SiNC)$\ga$240.

{\it{SiH, SiH$_4$}}.- In our frequency range there are not transitions of
SiH, a molecule tentatively detected by \citet{Sch01} towards
Orion KL, so we cannot assess that detection. The other Silicon
molecule detected in the space
(in IRC+10216) is SiH$_4$ \citep{Gol84}, but this molecule
can be only observed at IR wavelengths.

{\it{ob-SiC$_3$, l-SiC3, Si$_3$, SiCCO, SiCCS, o-SiH$_2$,
o-H$_2$CSi, p-H$_2$CSi, mb-Si$_2$H$_2$, o-db-Si$_2$H$_2$}}.- 
These molecules have not been
detected in the space yet. Upper limits for their column density are
shown in Table \ref{tab_nondetected}.

\section{Discussion}
\label{sect_dis}


SiO is a key tracer of shocked emission. Many interferometric
observations show that thermal and maser SiO emission depicted the low
velocity outflow centered in source $I$ in Orion KL \citep{Bla96, 
Wri96, Beu05, Pla09, God09, Zap09}. In addition,
SiO traces many other molecular outflows in different sources 
\citep{Jim04, Gib07, Deb09, Zap09}.
\citet{Moo07} have not
found emission from SiO at the position of the hot molecular core 
in G34.26+0.15. This hot core does not have a central source but rather
it is externally heated, similar to the Orion compact
ridge, by shocks, ionization fronts and stellar winds. \citet{Moo07} pointed
out that the lack of SiO in this hot core rules out any significant 
role played by shocks in determining the hot core chemistry.
Observations of SiO in the L1448-mm outflow permit to distinguish
between the shock precursor and the postshock components \citep{Jim05};
they observed an enhancement in the abundances of SiO (and another
shock tracers) by one order of magnitude in the shock precursor 
component and three orders of magnitude in the postshock gas (leading to
the broadening of the line profiles), evidence of recent ejection
of SiO from grains \citep{Flo96}.




\subsection{Molecular abundances}
\label{sect_dis_abu}

Molecular abundances were derived using the H$_2$ column density
calculated by means of the C$^{18}$O column
density (1.5$\times$10$^{16}$, 1.5$\times$10$^{16}$,
1$\times$10$^{17}$, 5$\times$10$^{16}$ cm$^{-2}$, and 2$\times$10$^{17}$ cm$^{-2}$
for the extended ridge, compact ridge, plateau, high velocity plateau,
and hot core, respectively) and the isotopic
abundance $^{16}$O/$^{18}$O=250, both provided in Paper I,
assuming that CO is a robust tracer of
H$_2$ and therefore their abundance ratio is roughly constant,
ranging from CO/H$_2$
$\simeq$ 5$\times$10$^{-5}$ (for the ridge components) to
2$\times$10$^{-4}$ (for the hot core and the plateau). In spite of 
the large uncertainty in this calculation, we
include it as a more intuitive result for
the molecules described in the paper.
We obtained N(H$_2$) = 7.5$\times$10$^{22}$, 7.5$\times$10$^{22}$,
2.1$\times$10$^{23}$, 6.2$\times$10$^{22}$, and 4.2$\times$10$^{23}$ cm$^{-2}$ 
for the extended ridge, compact ridge, plateau, high velocity plateau,
and hot core, respectively; for the 15.5 km s$^{-1}$ component
we assume $N$(H$_2$)=1.0$\times$10$^{23}$ cm$^{-2}$ as an average 
value in Orion KL. In addition, 
we assume that the
H$_2$ column density spatially coincides with the emission from the
species considered.
Our estimated source average abundances for each Orion KL component 
are summarized in
Table \ref{tab_abun} (only available online), 
together with comparison values from other
authors (\citealt{Sut95}, \citealt{Per07}, and \citealt{Ziu88}). 
The differences between the abundances shown in Table \ref{tab_abun}
are mostly due to the different
H$_2$ column density considered, to the assumed cloud component of the
molecular emission and discrepancies in
the sizes of these components.
\subsection{On the origin of the SiO and SiS emission} 
\begin{figure*}
\includegraphics[angle=270,scale=.60]{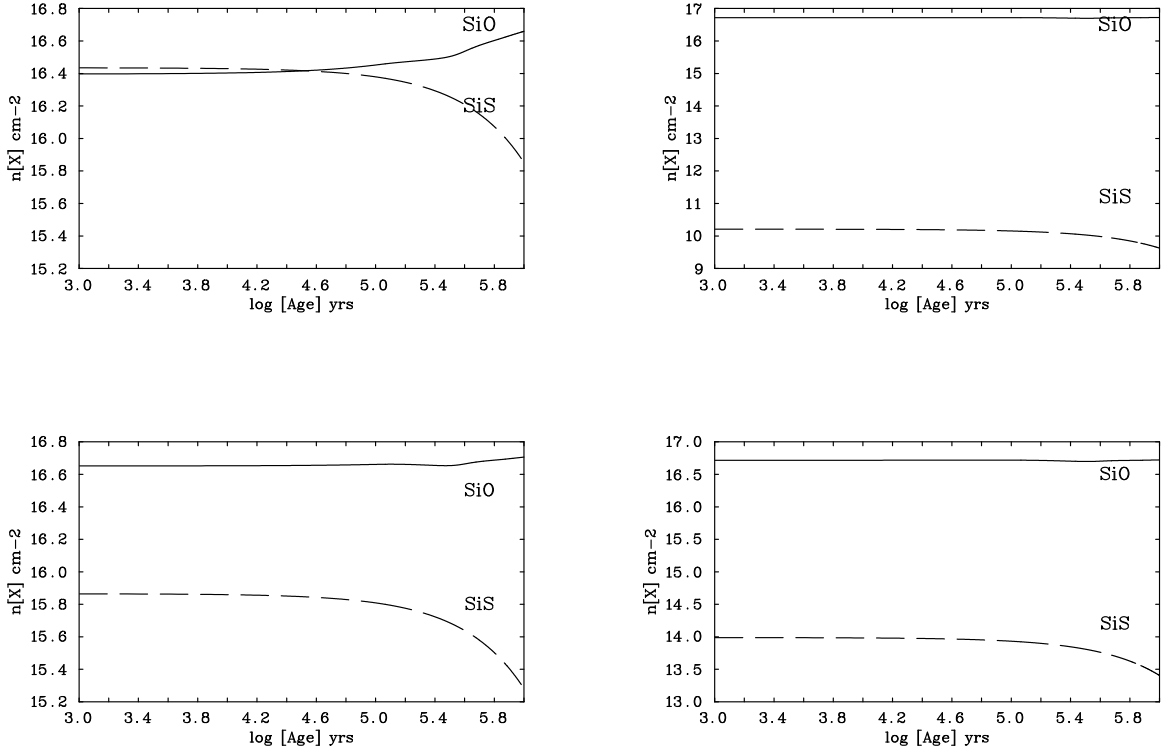}
\caption{Column densities of SiO (solid line) and SiS (dotted line) as a function of time 
from Phase II of the hot core model with an initial elemental abundance of 1x10$^{-6}$ for Sulphur and 
8x10$^{-8}$ for Silicon. The different plots show different efficiencies for the formation of 
SiS on the grain, namely: 100\% (top left); 0\% (top right); 30\% (bottom left); 10\% (bottom 
right).}
\label{fig_serena}
\end{figure*}

In order to qualitatively investigate the origin of the SiS and SiO emissions 
in Orion KL we ran a grid of models using the chemical model UCL\_CHEM 
\citep{Vit04a, Vit04b, Ler10}, a time and depth dependent gas-grain model. 

We modeled the hot core and the plateau separately. Both models are two phase calculation. 
In Phase I we follow the chemical and dynamical evolution of a collapsing core up 
to a final density of 5$\times$10$^7$ cm$^{-3}$ for the hot core component and 
10$^6$ cm$^{-3}$ for the plateau component as derived in Sect. \ref{sect_phy}.
The initial gas is at typical densities of $\sim$ 200 cm$^{-3}$ and in atomic form (apart 
from a fraction of hydrogen which is already in molecular form) and it 
undergoes a free-fall collapse \citep{Raw92} until the final 
densities are reached. During this time,
atoms and molecules from the gas freeze on to the dust grains and they 
hydrogenate where possible. Note that the advantage of this approach is that 
the ice composition is not assumed but it is derived by a time dependent computation
of the chemical evolution of the gas/dust interaction process. 
However, the ice composition does depend on the percentage of gas depleted on to the grains
during the collapse, and this in turns depend on the density as well as on the sticking 
coefficient and other properties of the species and of the grains (see \citealt{Raw92}).
In our model we can vary such percentage (reflecting the uncertainty on the grain 
properties and sticking probabilities) and the degree (or efficiency) of depletion (as well as the 
viability of different surface reactions) is explored in this study.
Our initial elemental abundances for Phase I are as
in \citet{Bel06} (see Table 1). We also ran some models where the initial 
abundances of both S and Si were either depleted or enhanced by a factor of 10 with respect to these values,
reflecting the uncertainty of their degree of depletion onto dust.
In Phase II, we follow the chemical evolution of the remnant core. 
For the hot core models, we simulate the effect of 
the presence of an infrared source in the center of the core or in its vicinity 
by subjecting the core to an increase in gas and dust temperature,
up to T = 300 K. This increasing of temperature is based on the luminosity of 
the protostar by using the observational luminosity function of \citet{Mol00}. 
The models we use have been published before: for the models describing the hot core
we refer the reader to \citet{Vit04a} and \citet{Ler10} while for the models describing the plateau we
refer the reader to \citet{Ler08}, \citet{Ler10} and \citet{Vit04b}.  
In both models the presence
of a non dissociative C-shock (modeled as in \citealt{Ber97}) can be simulated.
If a shock is included in the model then sputtering 
also occurs and is faster than thermal evaporation. 
We have ran a total of 6 models for the hot core component and 2 models for the 
plateau component. For the hot core model we investigated the sensitivity of the
chemical abundances to the degree of gas depleted on to the grains during the formation
of the core; the branching ratios of surface reactions relevant to the formation
of SiO and SiS; the initial abundances of Sulphur and Silicon; 
whether the gas is subjected to a non-dissociative shock during the hot core lifetime.
For the plateau models, we only varied the initial Sulphur and Silicon abundances.

We were able to reproduce the observed column density of SiO with most models.
The only constrain we found was that the temperature of the gas must be at least
$\sim$ 100 K or, alternatively, must have undergone a shock. SiS, on the other hand,
is difficult to produce: surface reactions (and subsequent evaporation or sputtering
of the mantles) seem to be necessary. We find that the only models
that succeed in reproducing the data are those where a percentage (even as small
as 5\%) of Sulphur on the grains react with Si to form SiS:
Figure \ref{fig_serena} shows the column density of SiO and SiS as a function of time during 
Phase II of the hot core for models differing only in the mantle formation 
efficiency of SiS (i.e on how efficient Si bonds with Sulfur). A qualitative match 
with the observation can be achieved, at early times, by 
those models where only 5\%-10\% of Sulphur on the grains react with Si to form SiS,
or at late times if a higher percentage of Sulphur is involved in the formation of SiS.

Note, however, that exact ages can not be derived from 
these considerations as the relationship of the 
time dependencies with the efficiency of SiS formation on grains will 
depend on the desorption times of SiO and SiS.

In conclusions, while it is possible to reproduce SiO in the gas phase (as well as
on the grains), our models indicate that SiS is a product of surface reactions,
most likely involving direct reactions of Sulphur with Silicon.

\section{Summary}
\label{sect_sum}

In Paper I we presented a line survey of 
Orion KL taken with the 30m IRAM telescope. The
sensitivity achieved allows to perform a line confusion
limited survey. 
Due to the wide frequency range covered and data 
quality we decided to present the line survey in a series of papers
focused in different molecular families. 
In this paper we presented the study of the emission from
Silicon-bearing species, SiO and SiS,
as well as their isotopologues and their vibrationally excited states.

For the $v$=1 state of SiO we have detected the $J$=2-1
line and, for the first time in this source, emission in the 
$J$=4-3 transition, both
of them showing strong masering effect.
The well known components of Orion (hot core, plateau, high velocity
plateau, extended ridge, and compact ridge) contribute to the observed
emission from SiO and its isotopologues whereas for SiS, $^{29}$SiS,
and SiS $v$=1 we have found emission from the hot core, the plateau,
and a feature at 15.5 km s$^{-1}$.
In order to check the origin of this feature we have observed the line
profiles of $J$=5-4 SiO around IRc2 and of different
molecules at an offset position (-15'', 15'') from IRc2.
We conclude that this extended feature  
is seen around the position 
$\Delta\alpha$=-7'', $\Delta\delta$=+7'' (near the BN object) with
a radius of $\simeq$5 arcseconds and probably
arises from the interaction of the outflow with
the ambient cloud.
  
The physical parameters obtained for each Orion KL component are
in agreement with those we can find in the Orion literature. For the
feature at v$_{LSR}$ = 15.5 km s$^{-1}$ we derive T$_K$=200 K, 
n(H$_2$)=5$\times$10$^6$ cm$^{-3}$, and v$_{half\;\;power\;\;intensity}$=7.5 km s$^{-1}$.
Column densities have been calculated with radiative transfer codes
based on either the LVG or the LTE approximations taking into account
the physical structure of the source and using the new SiO-p-H$_2$ 
and SiS-o-H$_2$ collisional rates. The results are provided as source 
averaged column densities. In this way, we obtain a total column
density of (7.4$\pm$2.0)$\times$10$^{15}$ and
(1.35$\pm$0.40)$\times$10$^{15}$ cm$^{-2}$ for SiO and SiS,
respectively.

We have derived several column density ratios which permit us to provide
the following average isotopic abundances: 
$^{28}$Si/$^{29}$Si=26$\pm$10, $^{29}$Si/$^{30}$Si=1.7$\pm$0.6, 
$^{16}$O/$^{18}$O$\ga$239, $^{18}$O/$^{17}$O$\simeq$2.
We have also investigated the origin of the SiS and SiO emission in Orion KL by the
use of a gas-grain chemical model and find that while SiO can be easily formed in the gas phase, SiS seems
to be a product of grain surface reactions, most likely involving 
direct reactions of Sulphur with Silicon.

The resulting vibrational temperature for SiS $v$=1 in the feature at
15.5 km s$^{-1}$ is $\simeq$ 500 K, larger than the kinetic
temperature derived for this component indicating an IR pumping or a
warmer component difficult to see in the lines of the ground
vibrational state.

Finally, we have derived upper limits for the column density of
non-detected molecules (Silicon-bearing species). For the detected 
species in other sources SiC, SiC$_2$,
c-SiC$_3$, SiC$_4$, SiN, SiCN, and SiNC the upper limits for their 
column density are 1.3$\times$10$^{14}$, 3.5$\times$10$^{13}$, 
1.3$\times$10$^{13}$, 3.5$\times$10$^{12}$,
6.1$\times$10$^{13}$, 3.1$\times$10$^{13}$, and 3.1$\times$10$^{13}$
cm$^{-2}$, respectively. 

\begin{acknowledgements}
We thank the Spanish MEC for funding support through grants
AYA2003-2785, AYA2006-14876, AYA2009-07304, ESP2004-665 and
AP2003-4619 (M. A.), Consolider project CSD2009-00038
the DGU of the Madrid Community government for support under IV-PRICIT
project S-0505/ESP-0237 (ASTROCAM). 
\end{acknowledgements}

\listofobjects

\Online
\begin{appendix}
\section{Density diagnostic}
In this section we study the influence of collisional rates on 
the modelling of the SiO and SiS lines observed in our line survey
of Orion. 
Although most of the emission in Orion arises from regions of
high volume density, the intensity of the high velocity wings apparent in the
SiO and SiS line profiles arise from a region, the plateau, in which
the density is not enough to thermalize the rotational levels of high
dipole moment molecules such as SiO (3.1 D) and SiS (1.73 D). Hence,
collisional rates are needed for each species to derive the physical
conditions. For SiO previous calculations by Green
and collaborators (\citealp{Bie83}; \citealp{Tur92}; see
  http://www.giss.nasa.gov/data/mcrates\#sio)
have provided collisional 
rates for this molecule and temperatures 20-1500 K.
These collisional rates have been recalculated using a new SiO-p-H$_2$ surface
by \citet{Day06}.
However, no collisional rates were available of SiS until recently:
\citet{Vin07} and \citet{Klo08} have calculated
the state to state collisional rates for the system SiS-He and SiS-H$_2$, 
respectively. 

We have tested the influence of collisional rates on 
the modelling of the SiO and SiS lines observed in our line survey
of Orion. 
Using the recent rates quoted above we have computed line
intensity ratios between some transitions 
to provide tools to estimate the physical conditions
in astrophysical sources. We have covered 
volumne densities between 10$^2$-10$^9$ cm$^{-3}$ and kinetic temperatures
from 10 to 300 K including levels up to $J$=20 and $J$=40 
for SiO and SiS, respectively. 
In the following we will analyse
the excitation conditions for 1$\to$0, 2$\to$1, 
3$\to$2, 5$\to$4, 12$\to$11, and 19$\to$18 for both molecules. 

\subsection{SiO}

\begin{figure*}
\includegraphics[angle=270,scale=0.7]{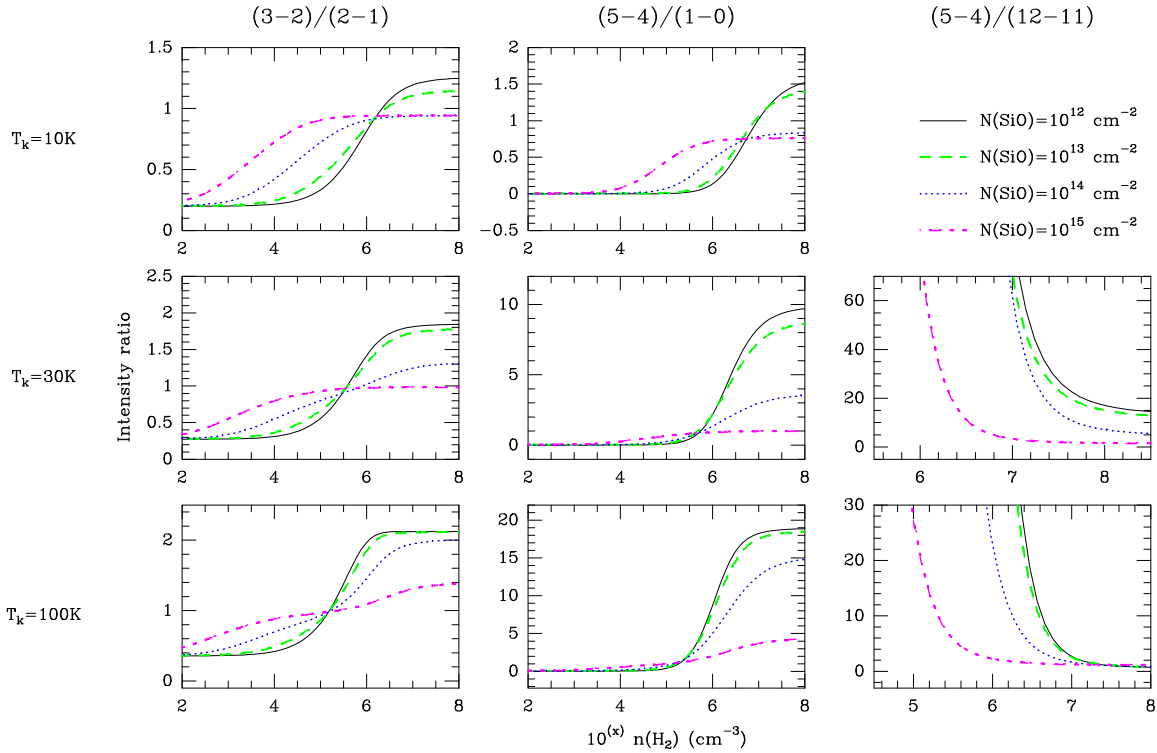}
\caption{Intensity ratio between two transitions for SiO as a function of 
H$_2$ density with the SiO column densities equal to 10$^{12}$, 10$^{13}$, 
10$^{14}$ and 10$^{15}$ cm$^{-2}$ and temperatures from 10 K to 100 K.}
\label{fig_sio_ratiodens}
\end{figure*}

\begin{figure*}
\includegraphics[angle=0,scale=0.8]{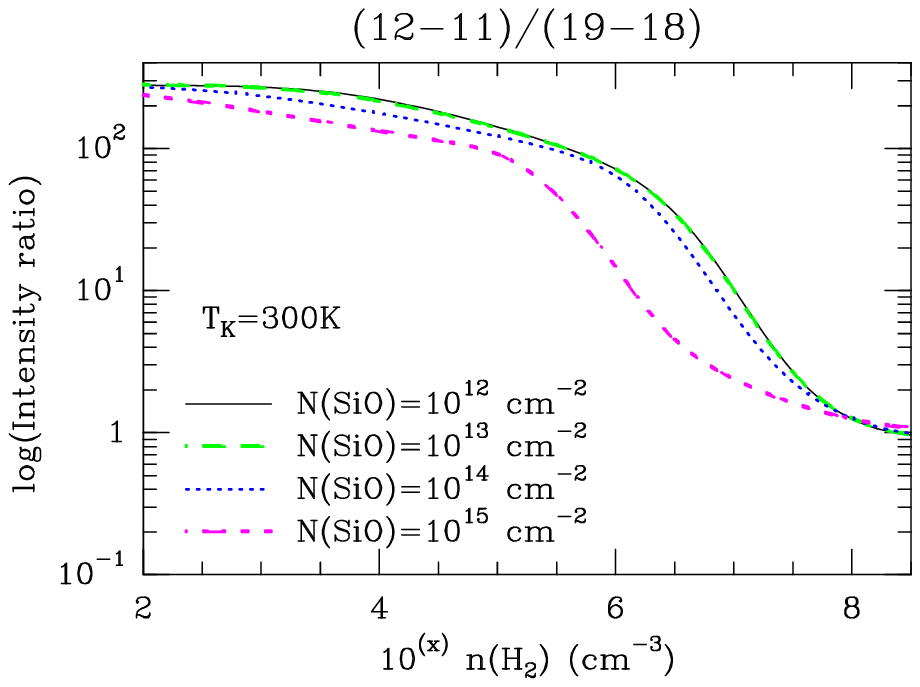}
\caption{Same as \ref{fig_sio_ratiodens} but for T = 300 K.}
\label{fig_sio_ratiodens2}
\end{figure*}

\begin{figure*}
\includegraphics[angle=270,scale=0.7]{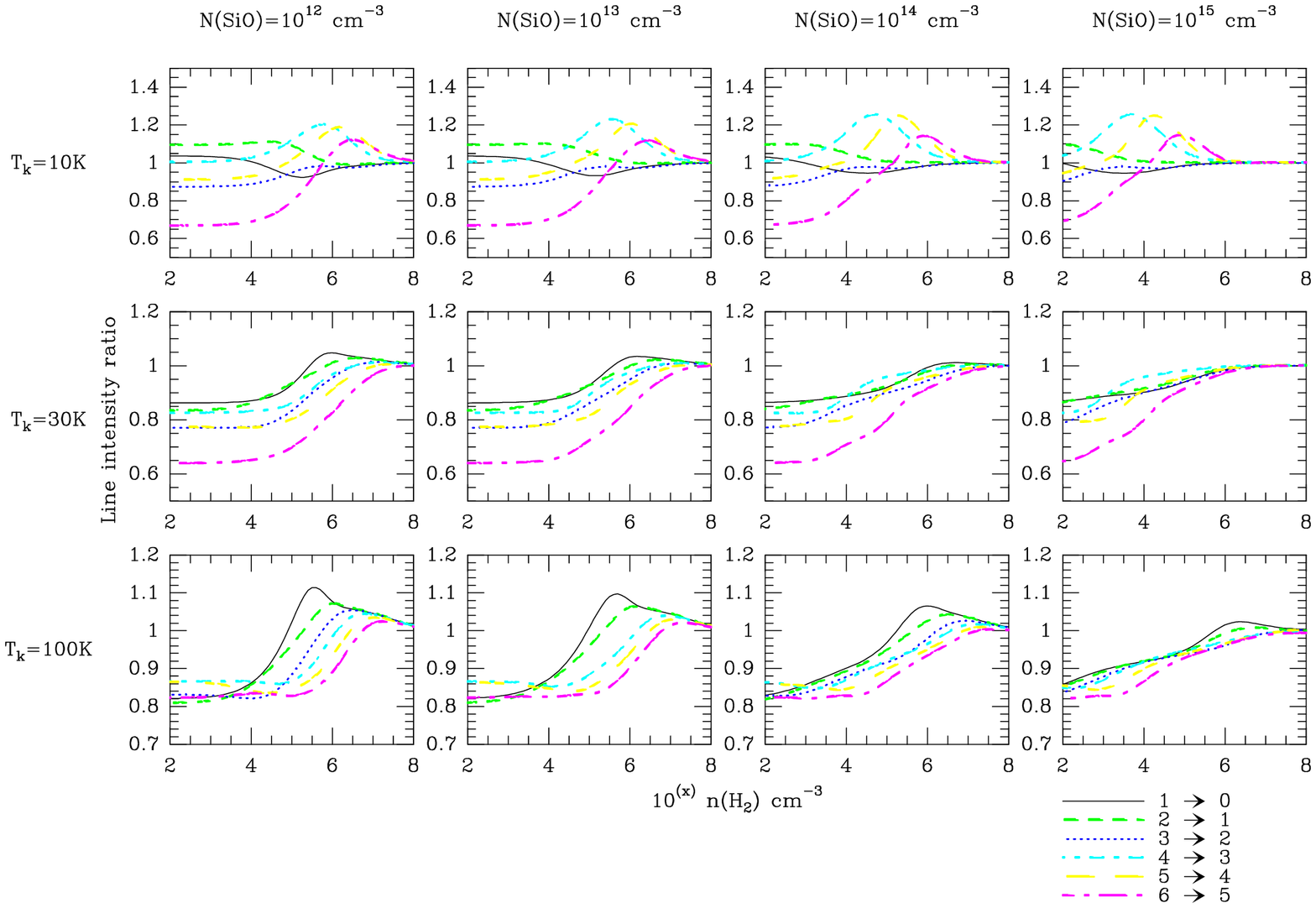}
\caption{Comparison of the line intensities predicted using two different sets of collisional rates
(see text) for different lines, column densities and densities.}
\label{fig_sio_rates}
\end{figure*}

The line intensity ratios for SiO are shown in Figs. \ref{fig_sio_ratiodens} 
(kinetic temperature between 10 and 100 K) and \ref{fig_sio_ratiodens2} (for a 
temperature of 300 K). 
In Fig. \ref{fig_sio_ratiodens} we can notice that the intensity ratio 
T$_B$(3$\to$2)/T$_B$(2$\to$1) is particularly sensitive to densities of 
about 10$^4$ to 10$^6$ cm$^{-3}$, whereas for T$_B$(5$\to$4)/T$_B$(1$\to$0) 
it shows important variations in the density range 10$^6$ to 10$^7$ cm$^{-3}$.
Using the intensity ratio T$_B$(5$\to$4)/T$_B$(12$\to$11), we can explore density values 
around 10$^7$-5$\times$10$^8$ cm$^{-3}$. For T$_K$=300 K, and T$_B$(12$\to$11)/T$_B$(19$\to$18),
we can trace densities above 10$^5$ cm$^{-3}$, as seen in Fig. \ref{fig_sio_ratiodens2}. 
For lower densities the 19-18 line will be very weak.

To illustrate this point, we deduce from our observations 
the intensity ratio T$_B$(2$\to$1)/T$_B$(3$\to$2) 
of SiO, and find $\simeq$2.6 (see Table 
\ref{tab_siolines}), which corresponds, depending on the column 
density and for T$_K$=100 K, to densities between 10$^{5.5}$ and
10$^7$ cm$^{-3}$. Our results for the physical parameters of the different
cloud components discussed in previous sections
are given in Table \ref{tab_prop}.

The collisional rates calculated by \citet{Tur92} have been used in the past
to describe collisions of SiO molecules with H$_2$. We have used through the
paper the new rates of \citet{Day06}.  
As a first step in
modelling SiO emission in warm clouds we have made a comparison between
the results using both sets of collisional rates for a large range of
physical conditions: T$_K$=10-300 K, N(SiO)=10$^{12}$-10$^{15}$ cm$^{-2}$,
and n(H$_2$)=1-10$^9$ cm$^{-3}$.
Figure \ref{fig_sio_rates} shows the brightness temperature ($T_B$)
ratio (predictions using the \citealp{Day06} rates over those obtained
from \citealp{Tur92} rates) as a function of H$_2$ density for 
different values 
of the SiO column density and temperature and for the six first transitions 
of this molecule. The plots in Fig. \ref{fig_sio_rates} show that the difference 
in the predicted
line intensities between both sets of collisional rates never exceed 40\%
($J$=6-5 line), being always below 20\% for all other rotational
lines and kinetic temperatures.
The lowest temperature for the
collisional rates of \citet{Tur92} is 20 K; we have extrapolated these rates to
obtain the corresponding ones at 10 K.
For most transitions the predicted line intensities 
from the \citet{Day06} rates are lower than
those predicted from \citet{Tur92} rates for densities below $\simeq$10$^5$
cm$^{-3}$.
Although the differences in line intensities
are not significant for the determination of the physical properties
of the emitting gas in interstellar clouds, we have adopted in our SiO calculations
the new rates of \citet{Day06}.

\subsection{SiS}

\begin{figure*}
\includegraphics[angle=270,scale=0.7]{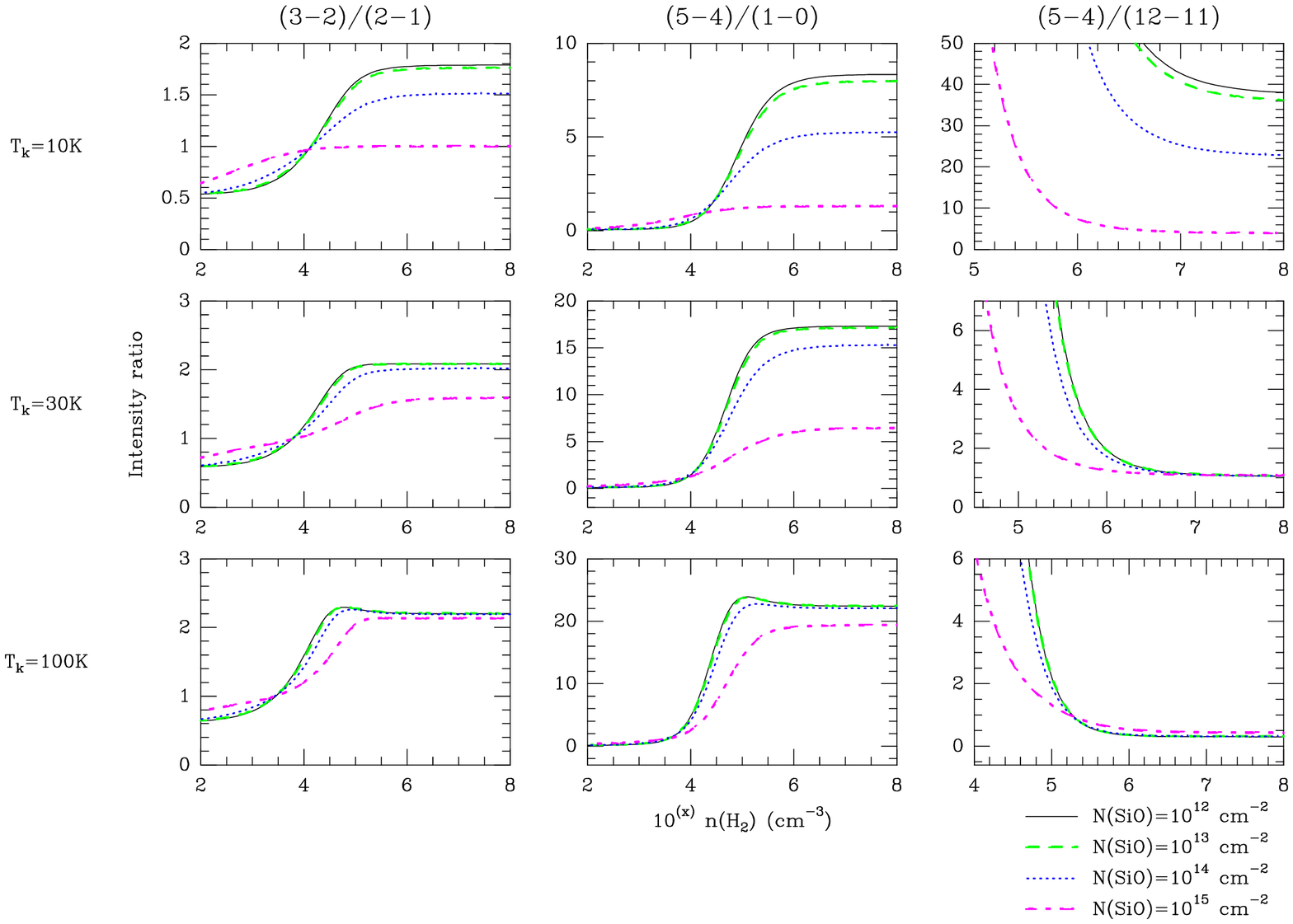}
\caption{Intensity ratio between two transitions for SiS as a function of 
H$_2$ density with the SiS column density equal to 10$^{12}$, 10$^{13}$, 
10$^{14}$ and 10$^{15}$ cm$^{-2}$.}
\label{fig_sis_ratiodens}
\end{figure*}

\begin{figure*}
\includegraphics[angle=0,scale=0.8]{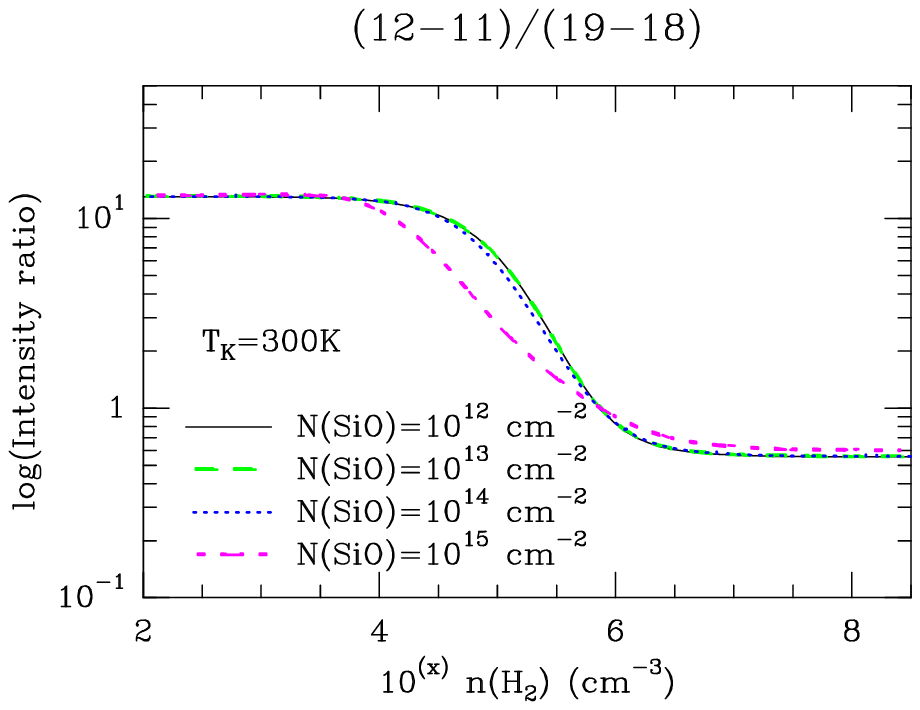}
\caption{Same as \ref{fig_sis_ratiodens} but for T = 300 K.}
\label{fig_sis_ratiodens2}
\end{figure*}

\begin{figure*}
\includegraphics[angle=270,scale=0.7]{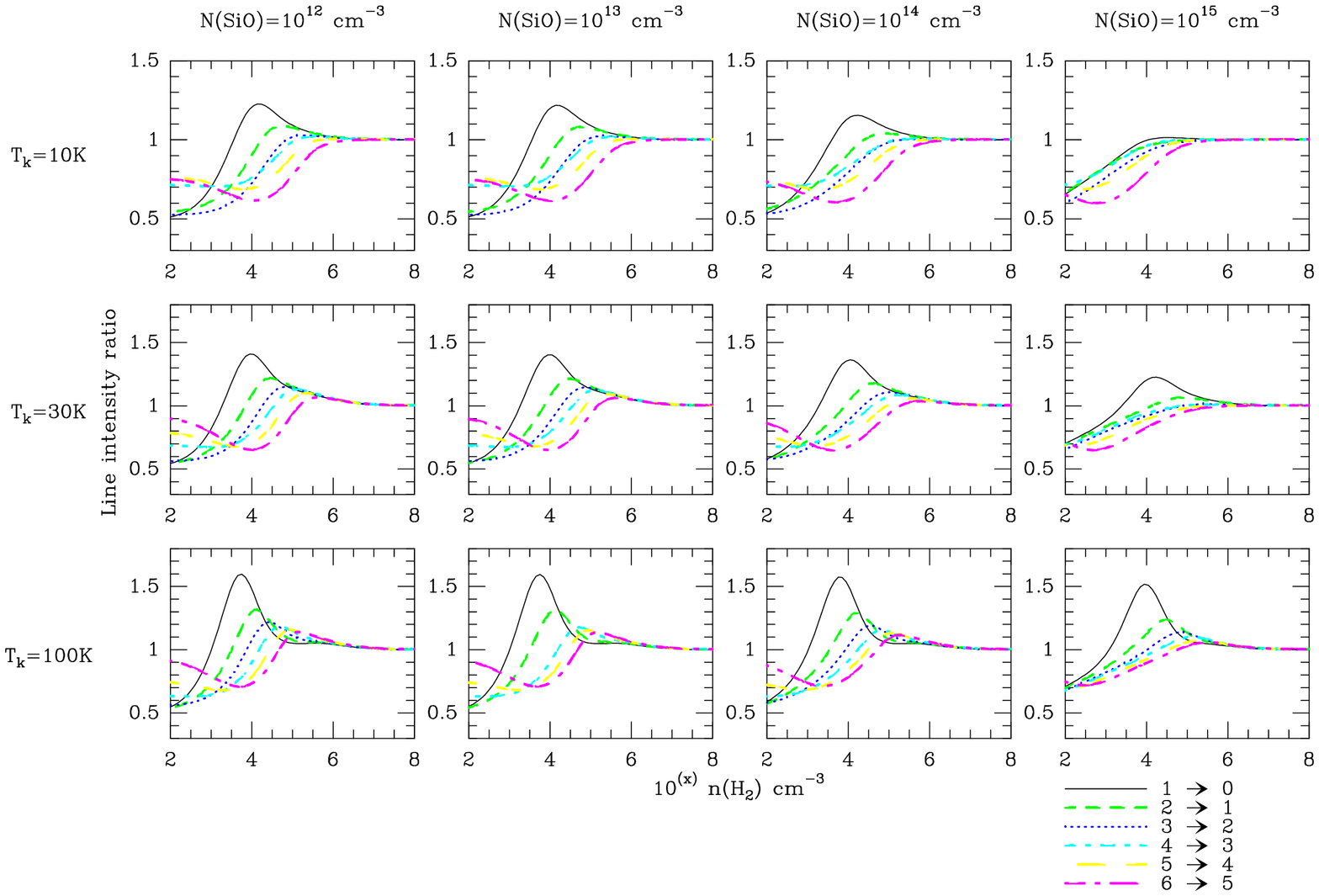}
\caption{Same as \ref{fig_sio_rates} for SiS with the SiS column density equal 
to 10$^{12}$, 10$^{13}$, 10$^{14}$ and 10$^{15}$ cm$^{-2}$.}
\label{fig_sis_rates}
\end{figure*}

We have performed the same study for SiS using the rates calculated
by \cite{Vin07}. Figs. 
\ref{fig_sis_ratiodens} and \ref{fig_sis_ratiodens2} show our results. We 
observe the same trends as for SiO, but for lower densities. Thus, the 
T$_B$(3$\to$2)/T$_B$(2$\to$1) ratio increases in the density range 10$^3$ to 10$^5$ 
cm$^{-3}$, the second one T$_B$(5$\to$4)/T$_B$(1$\to$0) in 10$^4$ to 
10$^5$ cm$^{-3}$, and the last graphs of Fig. \ref{fig_sis_ratiodens} 
show that 
T$_B$(5$\to$4)/T$_B$(12$\to$11) vary mostly between 10$^5$ and a few 10$^6$ cm$^{-3}$. In 
Fig. \ref{fig_sis_ratiodens2}, we can see that for T$_K$=300 K, 
T$_B$(12$\to$11)/T$_B$(19$\to$18) are useful for densities around 10$^5$ cm$^{-3}$.
Two of those lines have been detected in our survey, the transitions 5$\to$4 
and 12$\to$11, with an intensity ratio of $\simeq$0.12. For T$_K$=100\,K  
we can derive a volume density $\simeq$10$^{5.5}$ cm$^{-3}$.

Before 2007, no collisional data was available for SiS. Hence most authors have
adopted the SiO collisional rates for SiS. As quoted above, 
the system SiS-He has been studied by \citet{Vin07} and \citet{Liq08} have 
compared these rates with the correspondings for the SiS-H$_2$ system. 
They have 
shown that, when scaled by the square root of the collision reduced mass, the SiS-He
rates were a reasonable approximation to describe collisions with H$_2$. We have 
updated the LVG code by adding the new SiS-He and SiS-H$_2$ rates. We have 
calculated the line 
intensities and compared the results to those obtained with rates for SiO 
from \citet{Tur92} (multiplied by a factor 2 to take into account the larger 
geometrical size of the SiS molecule). Figure \ref{fig_sis_rates} shows the 
line intensity ratio (brightness temperature obtained with rates from 
\citealp{Vin07} over those calculated with rates from \citealp{Tur92}) 
as a function 
of n(H$_2$) for the six first rotational transitions.
The discrepancies in the line intensities between 
the two cases do not exceed 50\%. As for SiO, we can 
notice that the differencies between the two sets of collisional rates
are the same for all column 
densities: the line intensity predicted with rates for SiS are lower than 
those predicted from \citet{Tur92} rates for densities up to $\simeq$10$^3$. 
These differences are quite small and do not affect the determination of the 
physical parameters performed by authors that have used previously the
\citet{Tur92} rates in 
their models to interpret SiS observations. We tried also to describe SiS 
excitation by collisions with the rate coefficients for the system CS-He from 
\citet{Gre78} with an appropriate scaling factor. The predicted intensities
show a reasonable 
agreement with those derived from SiS-He or SiS-H$_2$ collisional
rates.
These results show that with the actual calibration accuracy for the SiS observations,
the determination of the properties of the emitting gas is not very 
sensitive to small differences in collisional rates. 
It is a rather curious result that the predicted line 
intensities with both sets of collisional rates, deduced from two different 
potential energy surfaces, moreover, with different propensity 
rules, produce similar intensity ratios (when the scaling factor is
adequately selected). In the determination of the SiS column densities we have
used the SiS-H$_2$ rates \citep{Klo08} although no significant differences are found when
we used the SiS-He ones.

In order to better quantify the effects of the use of scaled SiO-He 
rates in predicting SiS intensities, we have considered a case in which the lines 
are optically thin, 
for instance, n(H$_2$)=10$^5$ cm$^{-3}$, T$_K$=40 K and N(SiS)=3.10$^{12}$ 
cm$^{-2}$. The predicted line intensities from SiS-He rates are 
T$_B$(2$\to$1, 4$\to$3, 6$\to$5)=0.08, 0.26, 0.33 K. To obtain these results 
from scaled SiO-He rates we need a density of 7.10$^4$ cm$^{-3}$ or to reduce 
the scaling factor from 2 to $\simeq$1.5, i.e., an error lower than 2 in density.

We have compared predictions from the SiS-He rates with those obtained from the SiS-H$_2$
rates and we have found not significant variation between both. 
\end{appendix}

\clearpage

\begin{appendix}
\section{Online Figures}

\begin{figure*}
\includegraphics[angle=270,scale=.80]{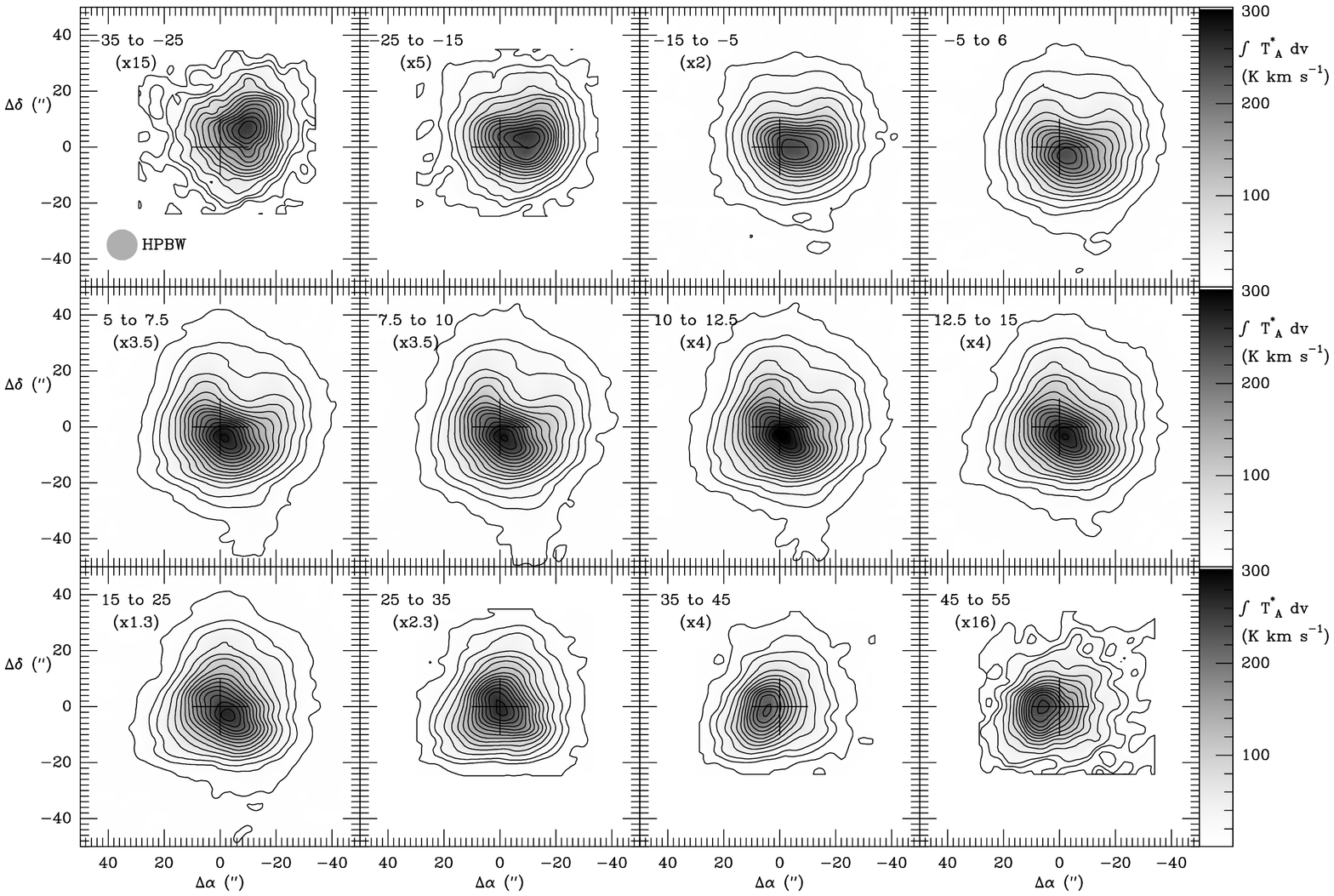}
\caption{Integrated intensity of the $J$=5-4
  transition of SiO at different velocity ranges (indicated at the top
  of each panel). In some panels, the integrated intensity of the maps
  has been multiplied by a scale factor (indicated in the panels) to
  maintain the same color dynamics for all maps. The step in
  integrated intensity ($\int$T$^*$$_A$dv) is
  20 K km s$^{-1}$ from 25 to 305 K km s$^{-1}$ and
  the minimum contours are 5 and 15 K km s$^{-1}$. See text
  (Sect. \ref{sect_obs}) for main beam antenna temperature correction.}
\label{fig_sio_map}
\end{figure*}

\begin{figure*}
\includegraphics[angle=0,scale=.75]{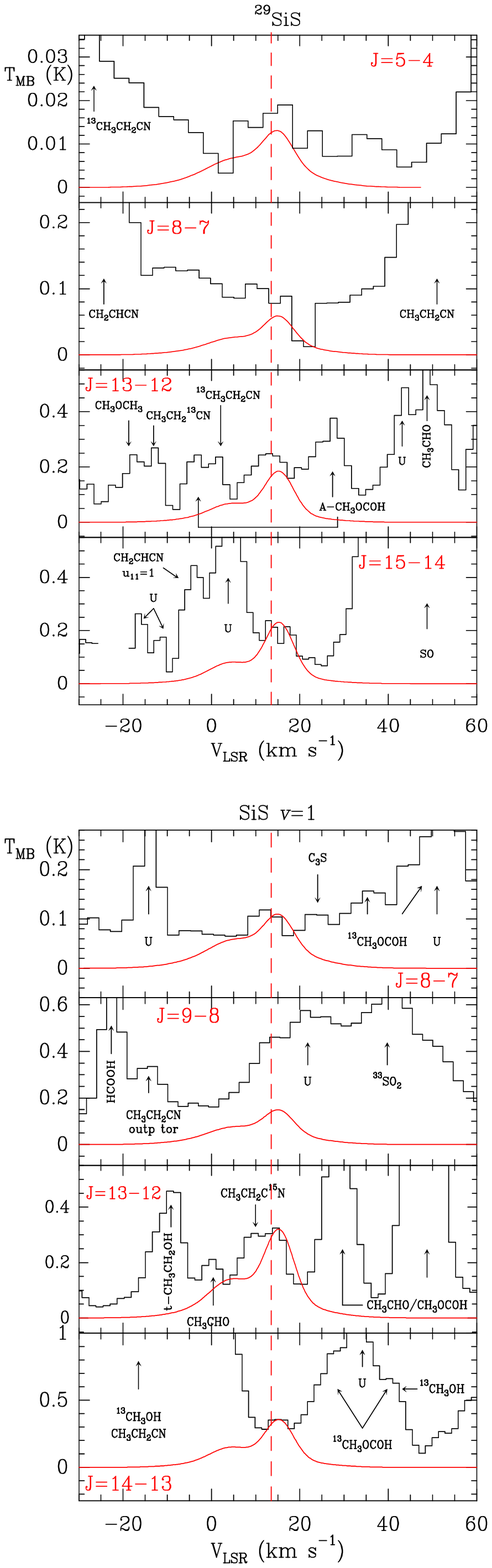}
\caption{Observed lines (histogram spectra) and model 
  (thin curves) of $^{29}$SiS and SiS $\textit{v}$=1. The dashed line
  shows a radial velocity at 13.5 km s$^{-1}$.}
\label{fig_29sis_sisv1}
\end{figure*}

\begin{figure*}
\includegraphics[angle=0,scale=1]{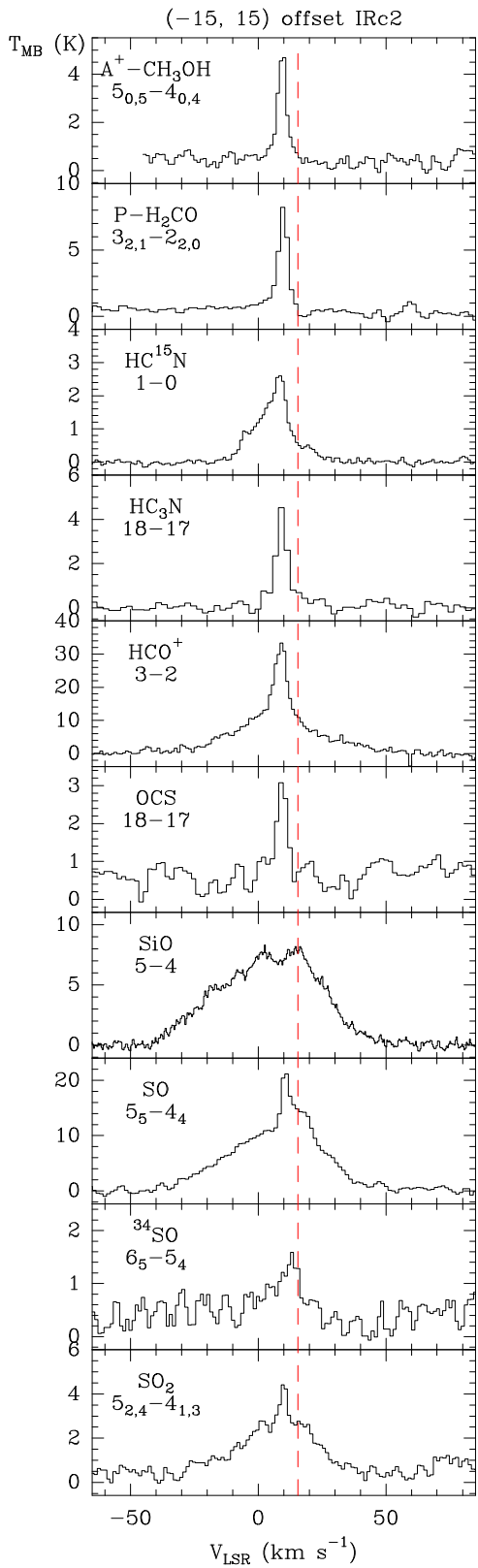}
\caption{Observed lines of different molecules at a position (-15'',
  15'') offset IRc2. The dashed line
  shows a radial velocity at 15.5 km s$^{-1}$.}
\label{fig_m15_15}
\end{figure*}

\begin{figure*}
\includegraphics[angle=0,scale=.90]{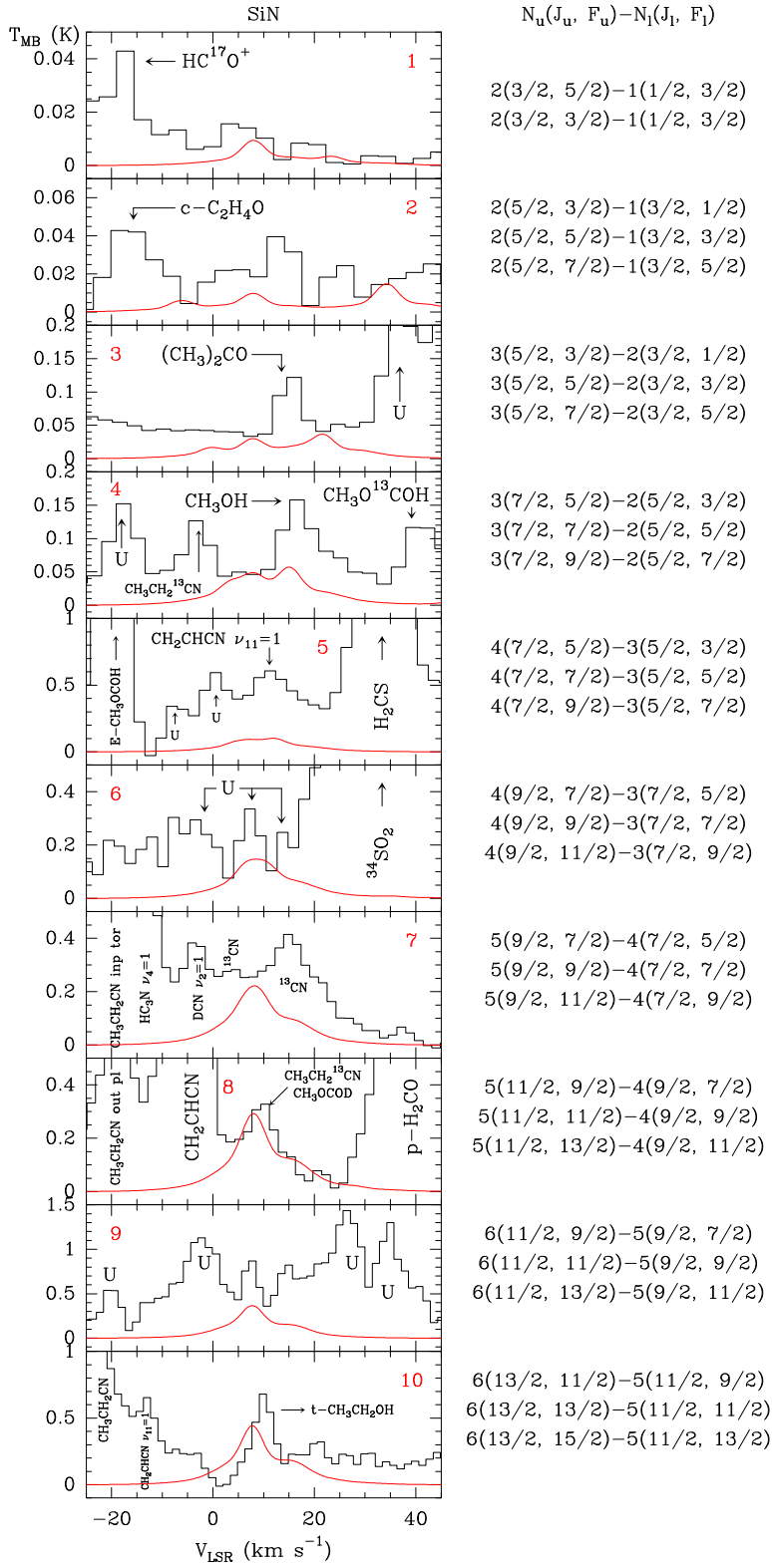}
\caption{Observed lines (histogram spectra) and model 
  (thin curves) of SiN.}
\label{fig_sin}
\end{figure*}

\end{appendix}

\clearpage

\begin{appendix}
\section{Online Tables}

\begin{table*}
\begin{center}
\caption{SiO and SiO isotopologues velocity components.\label{tab_sio_gau1}}
\begin{tabular}{lllllll}
\hline 
\hline
Species & \multicolumn{3}{c}{Wide component} &
\multicolumn{3}{c}{Narrow component} \\ 
 & v$_{LSR}$ (km s$^{-1}$) & $\Delta$v (km s$^{-1}$) & T$_{MB}$ (K) &
v$_{LSR}$ (km s$^{-1}$) & $\Delta$v (km s$^{-1}$) & T$_{MB}$ (K)\\
\hline
 SiO 2-1 & 9.4$\pm$0.5 & 44$\pm$2 & 5.19 & 8.45$\pm$0.13 & 19.1$\pm$0.5 & 13.3\\
 SiO 3-2 & 10.2$\pm$0.3 & 40.1$\pm$0.9 & 15.5 & 8.4$\pm$0.2 & 14.9$\pm$0.7 & 10.8\\
 SiO 4-3 & 12.0$\pm$0.3 & 41.7$\pm$0.9 & 26.6 & 8.5$\pm$0.2 & 15.8$\pm$0.8 & 18.8\\
 SiO 5-4 & 11.91$\pm$0.13 & 40.4$\pm$0.4 & 27.1 & 7.30$\pm$0.11 & 14.8$\pm$0.4 & 18.4\\
 SiO 6-5 & 9.7$\pm$0.3 & 40.8$\pm$0.5 & 46.1 & 7.7$\pm$0.3 & 13.3$\pm$0.8 & 15.6\\
 $^{29}$SiO 4-3 & 7.9$\pm$1.1 & 46$\pm$6 & 2.88 & 9.5$\pm$0.3 & 15.5$\pm$1.3 & 5.85\\
 $^{30}$SiO 4-3 & ... & ... & ... & 8.8$\pm$0.2 & 19.7$\pm$0.5 & 5.15\\
 $^{30}$SiO 5-4 & ... & ... & ... & 10.23$\pm$0.15 & 21.0$\pm$0.6 & 6.19\\
 $^{30}$SiO 6-5 & 7.8$\pm$0.5 & 46$\pm$2 & 4.35 & 8.6$\pm$0.2 & 16.8$\pm$0.7 & 8.50\\
\hline
\end{tabular}
\end{center}
\tablefoot{
Velocity components of selected SiO and SiO isotopologues lines derived
from two Gaussian fits.
}
\end{table*}

\begin{table*}
\begin{center}
\caption{SiS velocity components.\label{tab_sis_gau1}}
\begin{tabular}{lllllll}
\hline 
\hline
Species & \multicolumn{3}{c}{Wide component} &
\multicolumn{3}{c}{Narrow component} \\ 
 & v$_{LSR}$ (km s$^{-1}$) & $\Delta$v (km s$^{-1}$) & T$_{MB}$ (K) &
v$_{LSR}$ (km s$^{-1}$) & $\Delta$v (km s$^{-1}$) & T$_{MB}$ (K)\\
\hline
 SiS 6-5 & 8.8$\pm$0.5 & 24.0$\pm$0.9 & 0.39 & 15.4$\pm$0.3 & 7.0$\pm$0.5 & 0.23\\
 SiS 8-7 & 9.6$\pm$0.4 & 24.9$\pm$1.0 & 1.03 & 15.7$\pm$0.5 & 6.3$\pm$1.0 & 0.49\\
 SiS 9-8 & 8.7$\pm$0.4 & 24.6$\pm$0.7 & 1.33 & 15.7$\pm$0.2 & 6.6$\pm$0.5 & 1.15\\
 SiS 15-14 & ... & ... & ... & 15.7$\pm$0.3 & 12.3$\pm$0.9 & 2.69\\
\hline
\end{tabular}
\end{center}
\tablefoot{
Velocity components of selected SiS lines derived from Gaussian fits.
}
\end{table*}

\begin{table*} 
\begin{center}
\caption{Molecular abundances\label{tab_abun}}
\begin{tabular}{llllll}
\hline
\hline 
 Region & Species & X (This work) & X\tablefootmark{7} & X\tablefootmark{8} & X\tablefootmark{9}\\
 & & ($\times$10$^{-8}$) & ($\times$10$^{-8}$) &
 ($\times$10$^{-8}$) & ($\times$10$^{-8}$)\\
\hline
Extended & SiO & 0.05 & $<$0.05 & ... & ...\\
ridge\tablefootmark{1} & SiS & ... & ... & ... & ...\\
\hline
Compact & SiO & 0.20 & 0.80 & ... & ...\\
ridge\tablefootmark{2} & SiS & ... & ... & ... & ...\\
\hline
Plateau\tablefootmark{3} & SiO & 2.20 & 0.80 & 1.10 & 23.5\\
 & SiS & 0.30 & ... & ... & 0.40\\
\hline
High velocity & SiO & 2.20 & ... & 4.50 & ...\\
plateau\tablefootmark{4} & SiS & ... & ... & ... & ...\\
\hline
Hot & SiO & 0.20 & 0.60 & ... & ...\\
core\tablefootmark{5} & SiS & 0.30 & ... & ... & ...\\
\hline
15.5 km s$^{-1}$ & SiO & 1.00 & ... & ... & ...\\
component\tablefootmark{6} & SiS & 0.70 & ... & ... & ...\\
\hline 
\end{tabular}
\end{center}
\tablefoot{
Derived molecular abundances and comparation with other works.\\
\tablefoottext{1}{Assuming $N$(H$_2$)=7.5$\times$10$^{22}$ cm$^{-2}$.}
\tablefoottext{2}{Assuming $N$(H$_2$)=7.5$\times$10$^{22}$ cm$^{-2}$.}
\tablefoottext{3}{Assuming $N$(H$_2$)=2.1$\times$10$^{23}$ cm$^{-2}$.}
\tablefoottext{4}{Assuming $N$(H$_2$)=6.2$\times$10$^{22}$ cm$^{-2}$.}
\tablefoottext{5}{Assuming $N$(H$_2$)=4.2$\times$10$^{23}$ cm$^{-2}$.}
\tablefoottext{6}{Assuming $N$(H$_2$)=1.0$\times$10$^{23}$ cm$^{-2}$.}
\tablefoottext{7}{From \citet{Sut95}.}
\tablefoottext{8}{From \citet{Per07}.}
\tablefoottext{9}{From \citet{Ziu88}.}
}
\end{table*}

\end{appendix}

\begin{thebibliography}{}

\bibitem[Alberts et al., 1990]{Alb90}
Alberts, I. L., Grev, R. S. \& Schaefer III, H. F.
1990, \jcp, 93, 5046

\bibitem[Anders \& Grevesse, 1989]{And89}
Anders, E. \& Grevesse, N. 1989 GeCoA, 53, 197

\bibitem[Andreazza \& Marinho, 2007]{And07}
Andreazza, C. M. and Marinho, E. P. 2007, MNRAS, 380, 365

\bibitem[Apponi et al., 1999a]{App99a}
Apponi, A. J., McCarthy, M. C., Gottlieb, C. A. \& Thaddeus, P.
1999 \apj, 516 103

\bibitem[Apponi et al., 1999b]{App99b}
Apponi, A. J., McCarthy, M. C., Gottlieb, C. A. \& Thaddeus, P. 
1999, \jcp, 111, 3911

\bibitem[Apponi et al., 2000]{App00}
Apponi, A. J., McCarthy, M. C., Gottlieb, C. A. \& Thaddeus, P. 
2000, \apj, 536L, 55

\bibitem[Bell et al., 2006]{Bel06}
%
Bell, T. A., Roueff, E., Viti, S., Williams, D. A.
2006, MNRAS, 371, 1865

\bibitem[Becklin \& Neugebauer, 1967]{Bec67}
Becklin, E. E. \& Neugebauer, G. 
1967, \apj, 147, 799

\bibitem[Bergin et al., 1997]{Ber97}
Bergin, E. A., Goldsmith, P. F., Snell, R. L., \& Langer, W. D. 
1997, \apj, 482, 285 

\bibitem[Beuther et al., 2004]{Beu04}
Beuther, H., Zhang, Q., Greenhill, L. J. et al.
2004, \apj, 616L, 31

\bibitem[Beuther et al., 2005]{Beu05}
Beuther, H., Zhang, Q., Greenhill, L. J., et al.
2005, \apj, 632, 355 

\bibitem[Beuther \& Nissen, 2008]{Beu08}
Beuther, H. \& Nissen, H. D. 2008 \apj, 679L, 121 

\bibitem[Bieniek \& Green, 1983]{Bie83}
Bieniek, R. J. and Green, S. 1983, ApJ, 265, L29

\bibitem[Bizzocchi et al., 2006]{Biz06}
Bizzocchi, L., Degli Esposti, C. \& Dore, L. 
2006, A\&A, 455, 1161

\bibitem[Blake et al., 1996]{Bla96}
Blake, G. A., Mundy, L. G., Carlstrom, J. E., Padin, S., Scott, S. L.,
Scoville, N. Z. and Woody, D. P. 1996,
\apj, 472, L49

\bibitem[Bogey et al., 1990]{Bog90}
Bogey, M., Demuynck, C. \& Destombes, J. L. 
1990, A\&A, 232L, 19

\bibitem[Bogey et al., 1994]{Bog94}
Bogey, M., Bolvin, H., Cordonnier, M., et al.
1994, \jcp, 100, 8614

\bibitem[Botschwina et al., 2002]{Bot02}
Botschwina, P., Sanz, M. E.,  McCarthy, M. C. \& Thaddeus, P.
2002, \jcp, 116, 10719

\bibitem[Botschwina, 2005]{Bot05}
Botschwina, P. 2005, JMoSt: THEOCHEM,
724, 95

\bibitem[Cernicharo, 1985]{Cer85}
Cernicharo, J. 1985. Internal IRAM report (Granada: IRAM)

\bibitem[Cernicharo et al., 1986]{Cer86}
Cernicharo, J., Kahane, C., G\'omez-Gonzalez, J. \& Gu\'elin, M. 
1986, A\&A, 167L, 9

\bibitem[Cernicharo et al., 1989]{Cer89}
Cernicharo, J., Gottlieb, C. A., Guelin, M., Thaddeus, P. \& Vrtilek, J. M. 
1989, \apj, 341L, 25

\bibitem[Cernicharo et al., 1990]{Cer90}
Cernicharo, J., Thum, C., Hein, H., John, D., Garcia, P., Mattioco, F., 1990,
A.\&A., 231, L15

\bibitem[Cernicharo et al., 1991a]{Cer91a}
Cernicharo, J., Gu\'elin, M., Kahane, C., Bogey, M. \& Demuynck, C. 
1991, A\&A, 246, 213

\bibitem[Cernicharo et al., 1991b]{Cer91b}
Cernicharo, J., Bujarrabal, V. \& Lucas, R. 
1991, A\&A, 249L, 27

\bibitem[Cernicharo et al., 1994]{Cer94}
Cernicharo, J., Gonz\'alez-Alfonso, E., Alcolea, J., Bachiller, R., 
John, D., 1994, \apj, 432, L59

\bibitem[Cernicharo et al., 2000]{Cer00}
Cernicharo, J., Gu\'elin, M., Kahane, C., 2000, A\&A Supl. Series,
142, 181

\bibitem[Churchwell et al., 1987]{Chu87}
Churchwell, E., Felli, M., Wood, D.O.D., Massi, M-, 1987, \apj, 321, 516

\bibitem[Comito et al., 2005]{Com05}
Comito, C., Schilke, P., Philips, T. G., Lis, D. C., Motte, F. and
Mehringer, D. 2005 \apjs, 165, 127 

\bibitem[Cordonnier et al., 1992]{Cor92}
Cordonnier, M., Bogey, M., Demuynck, C. \& Destombes, J.-L. 
1992, \jcp, 97, 7984

\bibitem[Dayou \& Balan\c ca, 2006]{Day06}
Dayou, F. and Balan\c ca, C. 2006, A\&A, 459, 297

\bibitem[De Buizer et al., 2009]{Deb09}
De Buizer, J. M., Redman, R. O., Longmore, S. N., Caswell, J. \& Feldman, P. A. 
2009, A\&A, 493, 127

\bibitem[de Vicente et al., 2002]{deV02}
de Vicente, P., Mart\'in-Pintado, J., Neri, R. \& Rodr\'iguez-Franco, A. 
2002 \apj, 574L, 163

\bibitem[Dickinson \& Rodr\'iguez-Kuiper, 1981]{Dic81}
Dickinson, D. F. \& Rodr\'iguez-Kuiper, E. 1981, \apj, 247, 112

\bibitem[Doeleman et al., 1999]{Doe99}
Doeleman, S.S., Lonsdale, C.J., Pelkey, S., 1999, ApJ, 510, L55

\bibitem[Doeleman et al., 2004]{Doe04}
Doeleman, S. S., Lonsdale, C. J.,Kondratko, P. T., Predmore, C. R. 2004, ApJ,
607, 361

\bibitem[Flower et al., 1996]{Flo96}
Flower, D. R., Pineau des Forets, G., Field, D. \& May, P. W. 
1996, MNRAS, 280, 447

\bibitem[Gabriel, 1993]{Gab93}
Gabriel, W. 1993, CP, 174, 45

\bibitem[Genzel et al., 1980]{Gen80}
Genzel, R., Downes, D., Schwartz, P.R., Spencer, J.H., Pankonin, V., 
\& Baars, J., 1980, \apj, 239, 519

\bibitem[Genzel \& Stuzki, 1989]{Gen89}
Genzel, R., \& Stuzki, J., 1989, \apj, 574, 258

\bibitem[Gezari et al., 1998]{Gez98}
Gezari, D.Y., Backman, D.E., \& Werner, M.W., 1998, \apj, 509, 283

\bibitem[Gibb et al., 2007]{Gib07}
Gibb, A. G., Davis, C. J. \& Moore, T. J. T. 
2007, MNRAS, 382, 1213

\bibitem[Goddi et al., 2009]{God09}
Goddi, C., Greenhill, L. J., Chandler, C. J., et al.
2009, \apj, 698, 1165

\bibitem[Golhaber \& Betz, 1984]{Gol84}
Goldhaber, D. M. \& Betz, A. L. 
1984, \apj, 279L, 55

\bibitem[Gonz\'alez-Alfonso et al., 1996]{Gon96}
Gonz\'alez-Alfonso, E., Alcolea, J. \& Cernicharo, J. 
1996, A\&A, 313L, 13

\bibitem[Gonz\'alez-Alfonso \& Cernicharo, 1997]{Gon97}
Gonz\'alez-Alfonso, E. \& Cernicharo, J. 
1997, A\&A, 322, 938

\bibitem[Gordon et al., 2000]{Gor00}
Gordon, V. D., Nathan, E. S., Apponi, A. J., et al.
2000, \jcp, 113, 5311

\bibitem[Gottlieb et al., 1989]{Got89}
Gottlieb, C. A., Vrtilek, J. M. \& Thaddeus, P. 
1989, \apj, 343L, 29

\bibitem[Green \& Chapman, 1978]{Gre78}
Green, S., and Chapman, S. 1978, ApJS, 37, 169

\bibitem[Greenhill et al., 1998]{Gre98}
Greenhill, L.J., Gwinn, C.R., Schwartz, C., Moran, J.M., Diamond,
P.J., 1998, Nature, 396, 650

\bibitem[Greenhill et al., 2004]{Gre04}
Greenhill, L. J., Gezari, D. Y., Danchi, W. C., Najita, J., Monnier, J. D., \&
Tuthill, P. G. 2004, \apj, 605, L57

\bibitem[Grev \& Schaefer III, 1992]{Gre92}
Grev, R. S. \& Schaefer III, H. F. 
1992, \jcp, 97, 7990 

\bibitem[Gu\'elin et al., 2000]{Gue00}
Gu\'elin, M., Muller, S., Cernicharo, J., et al.
2000, A\&A, 363L, 9

\bibitem[Gu\'elin et al., 2004]{Gue04}
Gu\'elin, M., Muller, S., Cernicharo, J., McCarthy, M. C. \& Thaddeus, P. 
2004, A\&A, 426L, 49

\bibitem[Hirota \& Ishikawa, 1999]{Hir99}
Hirota, E. \& Ishikawa, H. 
1999, \jcp, 110, 4254

\bibitem[Hoeft et al., 1969]{Hoe69}
Hoeft, J., Lovas, F. J., Tiemann, E. \& T\"orring, T. 
1969, ZNatA, 24, 1422

\bibitem[Izuha et al., 1996]{Izu96}
Izuha, M., Yamamoto, S. \& Saito, S. 
1996, \jcp, 105, 4923

\bibitem[Jim\'enez-Serra et al., 2004]{Jim04} 
Jim\'enez-Serra, I., Martí\'n-Pintado, J., Rodr\'iguez-Franco, A.
\& Marcelino, N. 
2004, \apj, 603L, 49 

\bibitem[Jim\'enez-Serra et al., 2005]{Jim05} 
Jim\'enez-Serra, I., Mart\'in-Pintado, J. \& Rodr\'iguez-Franco,
A. 2005 \apj, 627, L121

\bibitem[Johansson et al., 1984]{Joh84} 
Johansson, L. E. B., Andersson, C., Elld\'er, J., Friberg, P.,
Hjalmarson, \AA., H\"olglund, B., Irvine, W. M., Olofsson, H. \& Rydbeck,
G. 1984 A\&A, 130, 227

\bibitem[Kerkines \& Mavridis, 2005]{Ker05} 
Kerkines, I. S. K. \& Mavridis, A. 
2005, \jcp, 123, l4301

\bibitem[Kleinmann \& Low, 1967]{Kle67} 
Kleinmann, D. E. \& Low, F. J. 
1967, \apj, 149L, 1

\bibitem[Klos \& Lique, 2008]{Klo08} 
Klos, J. \& Lique, F., 2008, MNRAS, 390 239  

\bibitem[Langhoff \& Bauschlicher, 1990]{Lan90} 
Langhoff, S. R. \& Bauschlicher, C. W., Jr. 
1990, \jcp, 93, 42L 

\bibitem[Lerate et al., 2008]{Ler08} 
Lerate, M. R., Yates, J., Viti, S., et al.
2008, MNRAS, 387, 1660L 

\bibitem[Lerate et al., 2010]{Ler10}
Lerate, M. R., Yatesi J., Barlow, M. J., Viti, S., Swinyard, B. M.
2010, MNRAS, 406, 2445L

\bibitem[Lique et al., 2008]{Liq08}
Lique, F., Tobola, R., Klos, J., Feautrier, N., Spielfiedel, A., Vincent, L., 
Chalasi\'nski, G., and Alexander, M. H. 2008, A\&A, 478, 567

\bibitem[Matthews et al., 2007]{Mat07}
Matthews, L. D., Goddi, C., Greenhill, L. J., Chandler, C. J., Reid, M. J., \&
Humphreys, E. M. L. 2007, in IAU Symp. 242, Astrophysical Masers and
their Environments, ed. J. M. Chapman \& W. A. Baan (Dordrecht: Kluwer),
130

\bibitem[McCarthy et al., 1999]{McC99}
McCarthy, M. C., Apponi, A. J. \& Thaddeus, P. 
1999, \jcp, 111, 7175

\bibitem[McCarthy et al., 2000]{McC00}
McCarthy, M. C., Apponi, A. J., Gottlieb, C. A. \& Thaddeus, P. 
2000, \apj, 538, 766

\bibitem[McCarthy et al., 2001]{McC01}
McCarthy, M. C., Apponi, A. J., Gottlieb, C. A. \& Thaddeus, P. 
2001, \jcp, 115, 870

\bibitem[McCarthy \& Thaddeus, 2003a]{McC03a}
McCarthy, M. C. \& Thaddeus, P. 
2003, \apj, 592L, 91

\bibitem[McCarthy \& Thaddeus, 2003b]{McC03b}
McCarthy, M. C. \& Thaddeus, P. 
2003, JMoSp, 222, 248

\bibitem[Menten \& Reid, 1995]{Men95}
Menten, K.M., Reid, M.J., 1995, ApJ, 445, L157

\bibitem[Menten et al., 2007]{Men07}
Menten, K.M., Reid, M.J., Forbrich, J., Brunthaler, A., 2007, ApJ, 474, 515

\bibitem[Millar, 1980]{Mil80}
Millar, T., J. 1980 Ap\&SS 72, 509M

\bibitem[Molinari et al., 2000]{Mol00}
Molinari, S., Brand, J., Cesaroni, R., \& Palla, F. 
2000, A\&A, 355, 617

\bibitem[Mookerjea et al., 2007]{Moo07}
Mookerjea, B., Casper, E., Mundy, L. G. \& Looney, L. W., 2007, \apj, 659, 447



\bibitem[Ohishi et al., 1989]{Ohi89}
Ohishi, M., Kaifu, N., Kawaguchi, K., et al.
1989, \apj, 345L, 83

\bibitem[Olofsson et al., 1981]{Olo81}
Olofsson, H., Hjalmarson, A. \& Rydbeck, O. E. H. 
1981, A\&A, 100L, 30

\bibitem[Pardo et al., 2001a]{Par01a}
Pardo, J. R., Cernicharo, J. and Serabyn, E. 2001, IEEE Tras. Antennas and
Propagation, 49, 12

\bibitem[Pardo et al., 2001b]{Par01b}
Pardo, J. R., Cernicharo, J., Herpin, F., et al.
2001, \apj, 562, 799

\bibitem[Persson et al., 2007]{Per07}
Persson, C. M., Olofsson, A. O. H., Koning, N., et al.
2007, A\&A 476, 807

\bibitem[Plambeck et al., 1990]{Pla90}
Plambeck, R.L., Wright, M.C.H., Carlstrom, J.E., 1990, \apj, 348, L65

\bibitem[Plambeck et al., 1995]{Pla95}
Plambeck, R.L., Wright, M.C.H., Mundy, L.G., \& Looney, L.W., 1995,
\apj, 502, L75

\bibitem[Plambeck et al., 2003]{Pla03}
Plambeck, R.L., Wright, M.C.H., Rao, R., 2003, 
\apj, 594, 911

\bibitem[Plambeck et al., 2009]{Pla09}
Plambeck, R.L., Wright, M.C.H., Friedel, et al., 2009, \apj, 704, L25

\bibitem[Raymonda et al., 1970]{Ray70}
Raymonda, J. W., Muenter, J. S. \& Klemperer, W. A. 
1970, \jcp, 52, 3458

\bibitem[Rawlings et al., 1992]{Raw92}
Rawlings, J. M. C., Hartquist, T. W., Menten, K. M., \& Williams, D. A. 
1992, MNRAS, 255, 471 

\bibitem[Saito et al., 1983]{Sai83}
Saito, S., Endo, Y. \& Hirota, E.
1983, \jcp, 78, 6447

\bibitem[Sanz et al., 2003]{San03} 
Sanz, M. E., McCarthy, M. C. and Thaddeus, 2003, P. \jcp, 119, v22, 11715

\bibitem[Sanz et al., 2005]{San05} 
Sanz, M. E., Alonso, J. L., Blanco, S. Lesarri, A. \& López, J. C. 
2005, \apj, 621L, 157

\bibitem[Schilke et al., 1997]{Sch97}
Schilke, P., Groesbeck, T. D., Blake, G. A. \& Philips,
T. G. 1997 \apjs, 108, 301

\bibitem[Schilke et al., 2001]{Sch01}
Schilke, P., Benford, C. J., Hunter, T. R., Lis, D. C. \& Philips,
T. G. 2001 \apjs, 132, 281

\bibitem[Schilke et al., 2003]{Sch03}
Schilke, P., Leurini, S., Menten, K. M. \& Alcolea, J. 2003 A\&A, 412, 15

\bibitem[Schwartz et al., 1982]{Sch82}
Schwartz, P.R., Zuckerman, B., Bologna, J.M., 1982, \apj, 256, L55

\bibitem[Snyder \& Buhl, 1974]{Sny74}
Snyder, L.E., \& Buhl, D., 1974, \apj, 189, L31

\bibitem[Sobolev, 1958]{Sob58}
Sobolev, V. V. 1958, en \textit{$''$Theorical
  Astrophysics$''$}. ed. Ambartsumyan, Pergamon Press Ltd. London
Cap. 29

\bibitem[Sobolev, 1960]{Sob60}
Sobolev, V. V. 1960, en \textit{$''$Moving Envelopes of
  Stars$''$}. Hardvard University Press.

\bibitem[Suenram et al., 1989]{Sue89}
Suenram, R. D., Lovas, F. J. \& Matsumura, K. 
1989, \apj, 342L, 103

\bibitem[Sutton et al., 1985]{Sut85}
Sutton, E. C., Blake, G. A., Masson, C. R. \& Philips,
T. G. 1985 \apjs, 58, 341


\bibitem[Sutton et al., 1995]{Sut95}
Sutton, E. C., Peng, R., Danchi, W. C., Jaminet, P. A., Sandell, G. \&
Russell, P. G. 1995 \apjs, 97, 455  

\bibitem[Tercero et al., 2010]{Ter10}
Tercero, B., Pardo, J. R., Cernicharo, and Goicoechea, J. R. 
submitted to A.\&A.

\bibitem[Thaddeus et al., 1984]{Tha84}
Thaddeus, P., Cummins, S. E. \& Linke, R. A. 1984 \apj, 283, 45

\bibitem[Turner, 1991]{Tur91}
Turner, B. E. 1991 \apjs, 76, 617

\bibitem[Turner et al., 1992]{Tur92}
Turner, B. E., Chan, K.-W., Green, S., and Lubowich, D. A. 1992, ApJ, 399, 114

\bibitem[Vasiliev et al., 1997]{Vas97}
Vasiliev, I., \"Og\"ut, S. \& Chelikowsky, J. R. 
1997, PhRvL, 78, 4805

\bibitem[Vincent et al., 2007]{Vin07}
Vincent, L., Lique, F., Spielfiedel, A., and Feautrier, N. 2007, A\&A, 472, 
1037

\bibitem[Viti et al., 2004a]{Vit04a}
Viti, S., Codella, C., Benedettini, M., \& Bachiller, R. 
2004, MNRAS, 350, 1029 
 
\bibitem[Viti et al., 2004b]{Vit04b}
Viti, S., Collings, M. P., Dever, J. W., McCoustra, M. R. S. \& Williams, D. A. 
2004, MNRAS, 354, 1141

\bibitem[Wright et al., 1990]{Wri90}
Wright, M. C. H., Carlstrom, J. E., Plambeck, R. L. and Welch,
W. J. 1990 AJ, 99, 1299

\bibitem[Wright et al., 1995]{Wri95}
Wright, M. C. H., Plambeck, R. L., Mundy, L.G., Looney, L.W., 
1995 \apj, 455, L185

\bibitem[Wright et al., 1996]{Wri96}
Wright, M. C. H., Plambeck, R. L. and Wilner, D. J. 1996 \apj, 469, 216

\bibitem[Zapata et al., 2009]{Zap09}
Zapata, L. A., Menten, K. Reid, M. \& Beuther, H. 
2009, \apj, 691, 332 

\bibitem[Ziurys \& Friberg, 1987]{Ziu87}
Ziurys, L. M. and Friberg, P. 1987 \apj, 314, L49

\bibitem[Ziurys, 1988]{Ziu88}
Ziurys, L. M. 1988, \apj, 324, 544

\bibitem[Ziurys, 1991]{Ziu91}
Ziurys, L. M. 1991, \apj, 379, 260

\end{thebibliography}
\end{document}